\documentclass{article}

\newif\ifNotUseNipsStyle
\NotUseNipsStyletrue

\newif\ifOmitHistogramSmallB
\OmitHistogramSmallBtrue

\ifNotUseNipsStyle
\usepackage[a4paper,margin=1in,footskip=0.25in]{geometry} 
\else
 \usepackage[preprint]{nips_2018}
\fi

\usepackage[utf8]{inputenc} % allow utf-8 input
\usepackage[T1]{fontenc}    % use 8-bit T1 fonts
\usepackage{hyperref}       % hyperlinks
\usepackage{url}            % simple URL typesetting
\usepackage{booktabs}       % professional-quality tables
\usepackage{multirow}
\usepackage{amsfonts}       % blackboard math symbols
\usepackage{amsmath}
\usepackage{nicefrac}       % compact symbols for 1/2, etc.
\usepackage{microtype}      % microtypography
\usepackage{graphicx}		% display figures

\graphicspath{{fig/}}
\usepackage{algorithm}
\usepackage{algorithmic}
\usepackage{subcaption}
\usepackage{multirow}
\usepackage{wrapfig}

\usepackage{nkj,nkj_e}

\newcommand{\CTBatch}{CT Batch}
\newcommand{\CTInherit}{ICT Inherit}
\newcommand{\CTGreedy}{ICT Greedy}
\newcommand{\CTRandom}{ICT Basic}

\newcommand{\GMMfull}{greedy max-min }
\newcommand{\CTBatchFull}{batch }

\newcommand{\ttt}{\textstyle}

\newcommand{\kmaxmin}{$k$-max-min diversification problem}
\newcommand{\ridfull}{relative inverse diversity}
\newcommand{\rid}{rel. inv. diversity}

\newcommand{\cities}{\textit{Cities}}
\newcommand{\faces}{\textit{Faces}}
\newcommand{\mnist}{\textit{MNIST}}

\newif\ifSuppressMemo
\ifSuppressMemo
\newcommand{\memo}[1]{}
\else
\usepackage{color}
\newcommand{\memo}[1]{{\bf \textcolor{red}{[#1]}}}
\fi

\title{Tight Bound of Incremental Cover Trees  \\
for Dynamic Diversification}

\author{
{  Hannah Marienwald$^{1,2}$, Wikor Pronobis$^1$}\\
{
\ifNotUseNipsStyle
\else
\bf
\fi
Klaus-Robert M{\"u}ller$^{1,2,3,4}$ and Shinichi Nakajima$^{1,2,5}$} \\
\ifNotUseNipsStyle
$^1$TU Berlin, $^2$Berlin Big Data Center, $^3$Korea University, \\
$^4$MPI for Informatics, $^5$AIP, RIKEN\\
\else
$^1$TU Berlin, $^2$Berlin Big Data Center, $^3$Korea University, $^4$MPI for Informatics, $^5$AIP, RIKEN\\
\fi
\texttt{hannah.marienwald@campus.tu-berlin.de}\\
\texttt{\{hannah.marienwald@campus., klaus-robert.mueller@,}\\
\texttt{nakajima@\}tu-berlin.de }\\
\texttt{wiktor.pronobis@gmail.com }\\
}
\ifNotUseNipsStyle
\date{}
\fi

\begin{document}
% \nipsfinalcopy is no longer used

\maketitle

\begin{abstract}
Dynamic diversification---finding a set of data points with maximum diversity from a time-dependent sample pool---%
is an important task in recommender systems, web search, database search, and notification services,
 to avoid showing users duplicate or very similar items.
The \emph{incremental cover tree} (ICT) with high computational efficiency and flexibility
has been applied to this task, and shown good performance.  
Specifically, it was empirically observed that ICT typically provides a set with its diversity only marginally ($\sim 1/ 1.2$ times) worse than the \GMMfull (GMM) algorithm, the state-of-the-art method for \emph{static} diversification with its performance bound optimal for any polynomial time algorithm.
Nevertheless, the known performance bound for ICT is 4 times worse than this optimal bound.
With this paper, we aim to fill this very gap between theory and empirical observations.  For achieving this, we first analyze variants of ICT methods, and derive tighter performance bounds. We then investigate the gap between the obtained bound and empirical observations by using specially designed artificial data for which the optimal diversity is known.  Finally, we analyze the tightness of the bound, and show that the bound cannot be further improved, i.e., this paper provides the {\em tightest} possible bound for ICT methods.
In addition, we demonstrate a new use of dynamic diversification for generative image samplers,
where 
prototypes are incrementally collected from a stream of artificial images generated by an image sampler.
\end{abstract}

\section{Introduction}
\label{sec:Introduction}

The diversification problem is a notoriously important issue in a number of popular applications, e.g.,  in recommender systems \cite{Ziegler05,Zhang08,Yu09}, web search \cite{Gollapudi09,Vieira11}, and database search \cite{Vee08,Demidova10,Liu09}. Diversity helps 
 to avoid showing users duplicate or very similar items, etc.
Formally, it is defined as follows:
\begin{definition} (Diversification problem)
Let $V = \{v_1, \ldots v_N\}$ be a set of $N$ points in some metric space $(\mathcal{V}, d)$ with distance metric $d: \mathcal{V} \times \mathcal{V} \rightarrow \mathbb{R}^+$. The goal of the \kmaxmin{} (for $k \leq N$) is to select a subset $S$ of $V$ such that
\begin{equation}\label{eq:DiversityMaximization}
\ttt
S = \argmax_{S' \subseteq V, |S'| = k} \; \; div(S'),
\end{equation}
where the diversity is defined as the minimum distance of any pair of data points in a set, i.e.,
\begin{equation}
\ttt
div(S') =\min_{s, s' \in S', s \neq s'} \; \; d( s, s' ).
\end{equation}
\end{definition}

This problem is known to be NP-hard \cite{Erkut90},
and variety of approximation methods have been proposed \cite{Erkut90,ravi1991facility,Erkut94,Vieira11,Minack11,drosou2014diverse}.
Among them, \GMMfull \cite{ravi1991facility} is one of the state-of-the-art methods, although it was proposed several decades ago \cite{Erkut94,drosou2014diverse}.
It was shown that GMM is guaranteed to provide an approximate solution with its diversity no smaller than $d^* / 2$,
where $d^*$ is the diversity of the optimal solution \cite{ravi1991facility}.
Furthermore, it was also shown that $d^* / 2$ is the best achievable bound by any polynomial time algorithm  \cite{ravi1991facility}.

\begin{figure}
  \centering
  \includegraphics[width=0.95\textwidth]{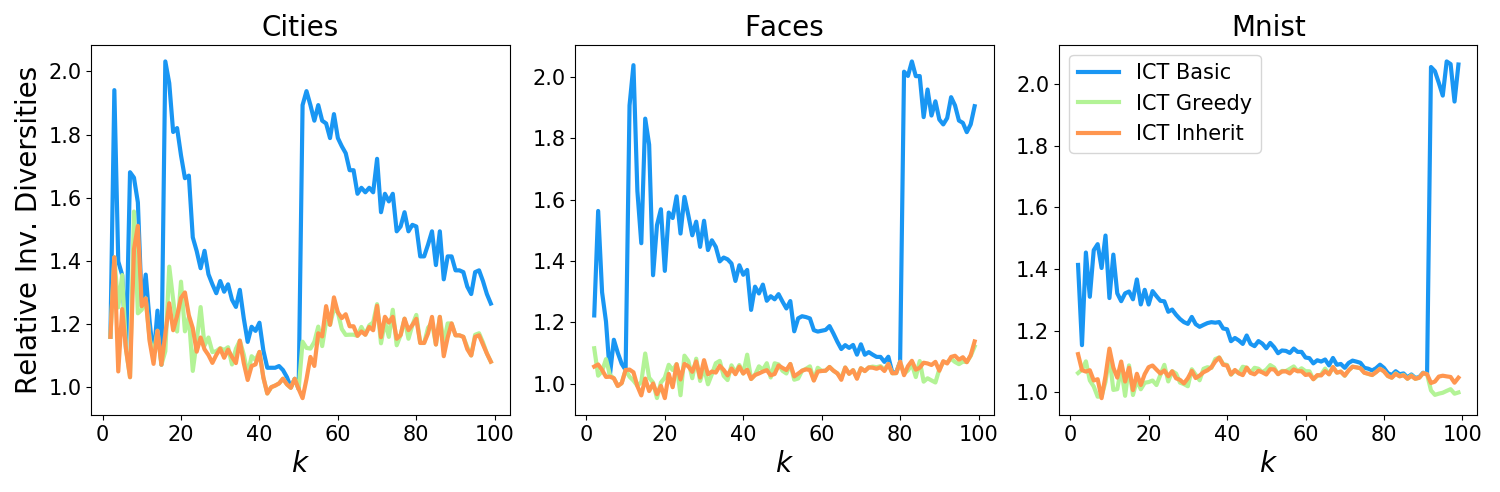}
  \caption{Achieved inverse diversities relative to GMM on the \cities, \faces{} and \mnist{} data set.  Detailed setting of experiments is described in Section~\ref{sec:Experiment}.}
\label{fig:Af_Ct_first}
\end{figure}

Recently, in many web applications including recommender systems, web search, and notification services,
the diversification algorithm is required to handle streaming data,
where
the sample pool $V$ is \emph{dynamic}---some new items can be added and some old items can be removed from time to time.
Drosou and Pitoura (2014) \cite{drosou2014diverse} argued that the cover tree (CT),
which was originally proposed for nearest neighbor search \cite{Beygelzimer06,Izbicki15}, is an appropriate tool to meet those streaming requirements.
In particular a family of CT approaches, called the \emph{incremental cover trees} (ICTs),
are suitable for {\em dynamic} diversification, because it allows addition and removal of items online.
The authors reported excellent empirical performance.  Especially, two variants, called \CTGreedy{} and \CTInherit,
typically provide a set with its diversity only $1/1.2$ times smaller than GMM
(see Fig.\ref{fig:Af_Ct_first} for our experimental results).

Nevertheless, the theoretically known performance bound of ICT is $d^*/8$ \cite{drosou2014diverse},
which is 4 times worse than GMM.
The goal of this paper is therefore to fill this gap between theory and empirical observations.
We theoretically analyze the property of ICT, and derive a tighter performance bound, $d^*/6$, 
for \CTGreedy{} and \CTInherit.
Then, we investigate the obtained bound and the empirical observations in details, by using specially designed artificial data for which the optimal diversity is known.
Finally, we analyze the tightness of performance bounds, and show that our proposed bounds cannot be improved further, i.e., for worst case analysis, this paper gives the tightest possible bounds of all ICT variants.
In addition, we demonstrate a new application of dynamic diversification for generative image samplers,
where 
prototypes are incrementally collected from a stream of artificial images generated by an image sampler.

\section{Background}
\label{sec:Background}

As discussed already above,
the \kmaxmin{} is known to be NP-hard \cite{ravi1991facility}. Therefore, various approximation algorithms have been proposed (see \cite{kunaver2017diversity} for an overview). 
In order to assess the accuracy of their approximation, a theoretical bound on the worst possible performance is usually given.
\begin{definition} \cite{williamson2011design} ($\alpha$-approximation algorithm) An $\alpha$-approximation algorithm for an optimization problem produces for all instances of the problem a solution whose value is within a factor $\alpha$ of the value of the optimal solution. 
$\alpha$ is called the approximation factor.
\end{definition}
Let $S^f = f(V, k)$ be a diverse subset of $V$ of size $k$, which was computed using method $f$. $f$ is  said to be an $\alpha$-approximation algorithm, iff for all possible diversification problems
\begin{equation} \label{eq:AF}
%div(S^f) \geq \frac{1}{\alpha} \cdot d^*,
div(S^f) \geq d^* / \alpha,
\end{equation}
where $d^*$ is the optimal diversity. 
Small values for $\alpha$ indicate better guaranteed performance, but because the approximate solution can at most be as good as the optimal solution, $\alpha \geq 1$.
Along with the computational complexity, the approximation factor is an important criterion that theoretically guarantees the performance of an algorithm.

\begin{minipage}[t]{.48\linewidth}
\begin{algorithm}[H]
\begin{algorithmic} [1]
\REQUIRE $V$ - a set of points, $ \; \; k$ - size of the subset
\ENSURE $S$ - diverse subset of $V$ of size $k$
\STATE $S \leftarrow $ randomly selected point $v \in V$
\WHILE{$\vert S \vert < k$}
\STATE $s^* \leftarrow \argmax_{v \in V \setminus S} \; \; \min_{s \in S} \; \; d(v,s)$
\STATE  $S \leftarrow S \cup \lbrace  s^* \rbrace$
\ENDWHILE
\RETURN $S$
\end{algorithmic}
\caption{Greedy Max-Min (GMM)}
\label{alg:GMM}
\end{algorithm}
\end{minipage}
\quad
\begin{minipage}[t]{.48\linewidth}
\begin{algorithm}[H]
\begin{algorithmic} [1]
\REQUIRE $CT$ - a cover tree for data set $V$, $ \; \; k$ - size of the subset
\ENSURE $S$ - diverse subset of $V$ of size $k$
\STATE $i \leftarrow i_{\max}$
\WHILE{$\vert C_i \vert < k$}
\STATE $i \leftarrow i - 1$
\ENDWHILE
\STATE $S \leftarrow$ select $k$ nodes from $C_i$
\RETURN $S$
\end{algorithmic}
\caption{Cover Tree for $k$-Max-Min Diversification Problem}
\label{alg:CT}
\end{algorithm}
\end{minipage}

In this section we summarize GMM, one of the most established algorithms, and cover tree-based approaches.
We give for each method the complexity and the approximation factor.

\subsection{Greedy Max-Min (GMM) Algorithm}\label{sec:gmm}
Although proposed several decades ago, \GMMfull (GMM) \cite{ravi1991facility} is still state-of-the-art for the diversification problem. A detailed description of the algorithm can be found in Algorithm \ref{alg:GMM}.
GMM approximates the diverse subset in a greedy manner, starting with either a randomly selected data point (line 1) \cite{abbar2013diverse} or the two most distant data points of the set \cite{ravi1991facility}. 
In subsequent iterations the data point with the largest pairwise distance to the current subset is added to the diverse set (line 3, 4). 
As the authors showed, GMM has an approximation factor $\alpha^{\mathrm{GMM}}$ of 2 and a complexity that is in $\mathcal{O}(N \cdot k)$.

\subsection{Incremental Cover-tree (ICT) Approaches}\label{sec:ctapproaches)}

Although, originally proposed for sublinear-time $k$-nearest neighbor search, the cover tree \cite{beygelzimer2006cover} can easily be adapted for diverse set approximation.
A cover tree for a data set $V$ is a leveled tree such that each layer of the tree covers the layer beneath it. 
Every layer of the tree is associated with an integer level $i$, which decreases as we descend the tree. 
The lowest layer of the tree holds the whole data set $V$ and is associated with the smallest level $i_{\min}$, whereas the root of the tree is associated with the largest level $i_{\max}$. 
A node in the tree corresponds to a single data point in $V$, but a data point might map to multiple nodes. 
However, any point can only appear once in each layer. Let layer $C_i$ be the set of all nodes at level $i$. 
For all level $i$ with $i_{\min} \leq i \leq i_{\max}$, the following invariants must be met: (1) 
Nesting: $C_i \subseteq C_{i-1}$. Once a point $p \in V$ appears in $C_i$, every lower layer in the tree has a node associated with $p$. (2) 
 Covering: For every $p \in C_{i-1}$, there exists a $q \in C_i$ such that $d(p,q) \leq b^i$ and the node in $C_i$ associated with $q$ is a parent of the node of $p$ in $C_{i-1}$.
(3) Separation:  For all distinct $p,q \in C_i, d(p,q) > b^i$.
Here $b > 1$ denotes the base of the cover tree. 
The cover tree was proposed with $b = 2$, but extended to arbitrary bases in \cite{drosou2014diverse} (cf.~Fig.~\ref{fig:covertree} in Appendix~\ref{sec:additionalfigures} for   cover tree examples).

Due to the invariants, each layer of the tree might already be an useful approximation. The general procedure of using a cover tree for the \kmaxmin, is shown in Algorithm \ref{alg:CT}. 
It aims to find the termination layer $C_i$ of the cover tree, i.e. the first layer that holds at least $k$ nodes.
In general $\vert C_i \vert \geq k$, thus, a subset of nodes must be selected (line 5). 
Possible selection strategies were introduced in \cite{drosou2014diverse} and are presented below.%
\footnote{
The authors of \cite{drosou2014diverse} also proposed a CT variant, called the cover-tree \CTBatchFull (\CTBatch),
with guaranteed approximation factor $\alpha = 2$.
 However, the cover-tree construction is as slow as GMM,
 and it cannot accept addition and removal of items in streaming data.
}

%\subsubsection{Cover-tree \CTRandomFull} 
\paragraph{\CTRandom} 
\CTRandom{} is the most straight-forward approach. 
It randomly selects $k$ nodes out of $C_i$, with its complexity in $\mathcal{O}(k)$. 
Because of the random selection, the diversity of the subset computed with \CTRandom{} might be the same as the diversity of the whole termination layer. 
%\begin{proposition}  \cite{Drosou12}
%\end{proposition}
%This methods works well in practical situations including streaming data setting.
%\subsubsection{Cover-tree \CTGreedyFull}
\paragraph{\CTGreedy}
\CTGreedy{} combines the cover tree approach with GMM. After the termination layer was located, we apply GMM on $C_i$ in order to select $k$ nodes.
Compared to the purely random approach, this selection strategy will, in most cases, give results with higher diversity.
By applying GMM only on $C_i$ instead of $V$, the complexity drastically reduces and is in $\mathcal{O}(\vert C_i\vert\cdot k)$.
%\subsubsection{Cover-tree \CTInheritFull}
\paragraph{\CTInherit}
\begin{wrapfigure}{r}{0.5\textwidth}
\begin{minipage}{1.0\linewidth}
\vspace{-7mm}
\begin{algorithm}[H]
\begin{algorithmic} [1]
\REQUIRE $CT$ - a cover tree for data set $V$, $ \; \; k$ - size of the subset
\ENSURE $S$ - diverse subset of $V$ of size $k$
\STATE $i \leftarrow i_{\max}$
\WHILE{$\vert C_i \vert < k$}
\STATE $i \leftarrow i - 1$
\ENDWHILE
\STATE $S \leftarrow C_{i+1}$
\WHILE{$\vert S \vert < k$}
\STATE $s^* \leftarrow \argmax_{c \; \in \; C_i \setminus \left( S \cup C_{i+1} \right)}  \min_{s \in S}\; d(c,s)$
\STATE  $S \leftarrow S \cup \lbrace  s^* \rbrace$
\ENDWHILE
\RETURN $S$
\end{algorithmic}
\caption{Cover Tree Inherit}
\label{alg:CTInherit}
\end{algorithm}
\end{minipage}
\end{wrapfigure}
\CTInherit{} is the most enhanced approach. 
It maintains the performance of \CTGreedy{} but further reduces the complexity. 
A detailed description can be found in Algorithm \ref{alg:CTInherit}. 
Instead of applying GMM on the whole layer $C_i$, we initialize the diverse subset with the previous layer $C_{i+1}$ (line 5) and only select some nodes from the termination layer (line 6 to 9).
Due to the separation invariant of the cover tree, $C_{i+1}$ already has a high diversity and, therefore, is an adequate initialization for the selection process.
The complexity of \CTInherit{} is in $\mathcal{O}\left(\vert C_i\setminus C_{i+1}\vert\cdot \left(k - \vert C_{i+1}\vert\right)\right)$.

\subsubsection{Approximation Factor}\label{sec:AFDrosou}
In \cite{drosou2014diverse} a first attempt to estimate the approximation factor was made.
\begin{proposition}\cite{drosou2014diverse} For $f \in \lbrace \mathrm{\CTRandom}, \mathrm{\CTGreedy}, \mathrm{\CTInherit} \rbrace$
\begin{equation}\label{eq:AFDrosou}
\ttt
\alpha^f = \frac{2b^2}{b-1}.
\end{equation}
\end{proposition}
For a cover tree built with $b=2$, this results in an approximation factor of $\alpha = 8$, regardless of which selection strategy is used.
However, as we will see in Sec. \ref{sec:Theory}, a lower and tighter approximation factor for \CTGreedy{} and \CTInherit{} can be proven.

\subsection{Diversification of Dynamic Data}
In many applications the set, for which a diverse subset is required, is not static. 
New data points must be added or old ones have to be removed from time to time.
Unfortunately, GMM is not an adequate choice for the diversification of dynamic data. Whenever the data set $V$ is changed, GMM has to be rerun from scratch.
Whereas the cover tree is a dynamic data structure. 
Insertion and removal of data points state no problem and the cover tree can easily be used for the diversification of dynamic data using the approaches presented above. Adding or removing a data point has complexity $\mathcal{O}(c^6 \cdot N \log N)$ where $c$ is the expansion constant \cite{beygelzimer2006cover, drosou2014diverse}.

\section{Theoretical Analysis}
\label{sec:Theory}
Because of the cover tree properties, each layer of the cover tree might already be an appropriate starting point for the approximation of the diverse set.
However, depending on the selection strategy, the quality of the diversity is likely to differ.
We expect \CTGreedy{} and \CTInherit{} to give results with higher diversity than \CTRandom.
In the following we will prove that the approximation factor derived in \cite{drosou2014diverse} is only a loose bound when it comes to the GMM-based selection strategies.
\begin{theorem}
 Let $C_i$ be the termination layer, i.e. the first layer of the cover tree that holds at least $k$ nodes
\begin{equation}\forall j > i,  |C_j| < k,\end{equation}
and $\beta \in \mathbb{R}^+$ be defined, such that
\begin{equation}\label{eq:beta}
\ttt
b^i = \frac{d^*}{\beta}.
\end{equation}
The diversity of every subset of $C_i$ can be bounded
\begin{equation} \label{eq:CTboundsep}
\forall \; C \subseteq C_i \, : \, {div}\left( C \right) \geq \frac{d^*}{\beta}. 
\end{equation}
Furthermore, $C_i$ will hold a subset $C$ of $k$ nodes with diversity of at least
\begin{equation}\label{eq:theorem1div}
\ttt
\exists \; C \subseteq C_i, \vert C \vert = k \,: \, {div}\left( C \right) \geq \max \left\lbrace  \frac{d^*}{\beta}, d^* (1 - \frac{1}{\beta}\frac{2b}{b-1})\right\rbrace 
\end{equation}
and $\beta < \frac{2b^2}{b-1}$. It follows
\begin{equation}\label{eq:CTboundalpha}
\ttt
\frac{d^*}{div(C)} \leq 1 + \frac{2b}{b-1}.
\end{equation}
\end{theorem}
%Eq. \eqref{eq:theorem1div} gives a lower bound on the diversity of the most diverse subset of $C_i$. 
%Eq. \eqref{eq:CTboundalpha} gives a limit on the relative diversity.
We give a sketch of the proof. For a detailed discussion see Appendix \ref{sec:prooftheorem1}.
Eq. \eqref{eq:CTboundsep} and the first part of the maximum in Eq. \eqref{eq:theorem1div} follows from the separation property of the cover tree. 
The latter part of the maximum follows from the covering and nesting property. Any layer of the cover tree covers the whole data set.
As a consequence $C_i$ is guaranteed to hold a subset of nodes, that cover the optimal diverse set.
The triangle inequality can then be used, to bound the distance between those nodes. This gives rise to the existence of $C$ and the latter part of the maximum. 
\QED

\begin{figure}
\centering
\begin{subfigure}{.4\textwidth}
  \centering
  \includegraphics[width=1.\linewidth]{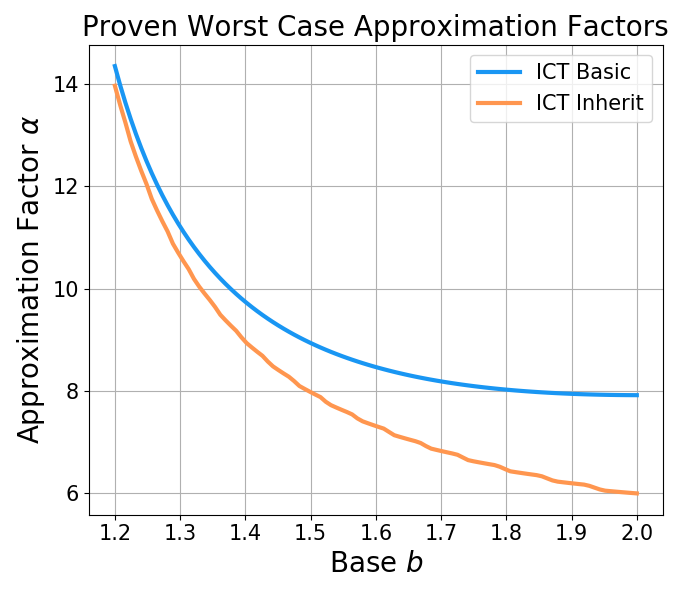}
\end{subfigure}%
\begin{subfigure}{.4\textwidth}
  \centering
  \includegraphics[width=1.\linewidth]{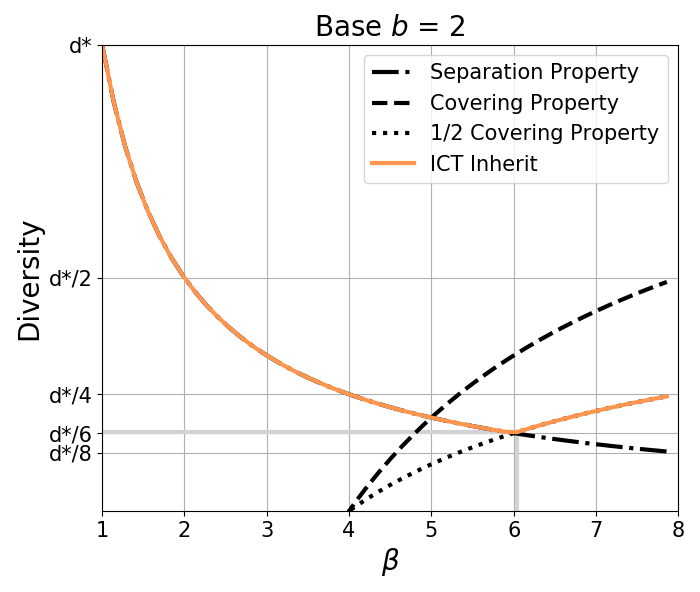}
\end{subfigure}
\caption{Visualization of the theoretical bounds derived in Sec. \ref{sec:Theory}. (left) Guaranteed approximation factor for \CTRandom{} and \CTInherit{} for different values of the base $b$.
  \CTGreedy{} has the same approximation factor as \CTInherit. (right) The bound on the diversity derived in Eq. \eqref{eq:theorem1div} for various values of $\beta$ and $b=2$. \textit{Separation Property} shows the first part of the maximum and \textit{Covering Property} shows the latter part of the maximum. The intercept of both lines shows what is stated in Eq. \eqref{eq:CTboundalpha}. 
  %\textit{CT Inherit} shows what is stated in Eq. \eqref{eq:ctinherit_div} (\CTInherit) and Eq. \eqref{eq:ctgreedydiv} (\CTGreedy). 
  The worst case approximation factor is 6.}
\label{fig:theory_rl}
\vspace{-3mm}
\end{figure}

For some $\beta$ it is crucial to select a subset of $C_i$ in an appropriate manner.
Thus, the diversity strongly depends on the strategy, that is used to select $k$ nodes out of $C_i$.

\begin{corollary}
Let $S^{\mathrm{\CTGreedy}} \subseteq C_i$ be a subset of $k$ nodes selected from $C_i$ using \emph{GMM} (\CTGreedy). The diversity of $S^{\mathrm{\CTGreedy}}$ is at least
\begin{equation}\label{eq:ctgreedydiv}
\ttt
div\left( S^{\mathrm{\CTGreedy}} \right) \geq \text{max} \left\lbrace  \frac{d^*}{\beta}, \frac{1}{2} \, d^* (1 - \frac{1}{\beta}\frac{2b}{b-1})  \right\rbrace.
\end{equation}
The approximation factor $\alpha$ of \CTGreedy{} is given by
\begin{equation}
\ttt
\frac{d^*}{div(S^{\mathrm{\CTGreedy}})} \leq 2 + \frac{2b}{b-1} = \alpha^{\mathrm{\CTGreedy}}.
\end{equation} 
\end{corollary}
For a detailed proof see Appendix \ref{sec:proofcorollarly1}.
The diversity of the selected subset cannot be worse than $\nicefrac{d^*}{\beta}$, because that bound is given by the separation criterion. 
As it was shown in \cite{ravi1991facility}, GMM has an approximation factor of $2$.
The best possible diversity in layer $C_i$ is given by Eq. \eqref{eq:theorem1div}. We get what is stated in Eq. \eqref{eq:ctgreedydiv}.
\QED

For $b=2$, this results in an approximation factor of $\alpha^{\mathrm{\CTGreedy}} = 6$. Compared to \CTGreedy, \CTInherit{} has a lower complexity.
Moreover, \CTInherit{} has the same bound on the diversity and the same approximation factor.
\begin{theorem}\label{theorem:ctinherit}
Let $S^{\mathrm{\CTInherit}} \subseteq C_i$ be a subset of $k$ nodes selected from $C_i$. It holds all nodes from the previous layer $C_{i+1}$ and remaining nodes were selected from $C_i\setminus C_{i+1}$ using GMM (\CTInherit). The bound of the diversity of $S^{\mathrm{\CTInherit}}$ is the same as the bound for $div\left( S^{\mathrm{\CTGreedy}} \right)$, i.e.
\begin{equation}\label{eq:ctinherit_div}
\ttt
div\left( S^{\mathrm{\CTInherit}} \right) \geq \max \left\lbrace  \frac{d^*}{\beta}, \frac{1}{2} \, d^* (1 - \frac{1}{\beta}\frac{2b}{b-1})  \right\rbrace.
\end{equation}
The approximation factor $\alpha$ of \CTInherit{} is given by
\begin{equation}
\ttt
\frac{d^*}{div(S^{\mathrm{\CTInherit}})} \leq 2 + \frac{2b}{b-1} = \alpha^{\mathrm{\CTInherit}}.
\end{equation} 
\end{theorem}
For a detailed proof, see Appendix \ref{sec:prooftheorem2}. Instead of starting with a randomly selected data point, \CTInherit{} initializes GMM with $C_{i+1}$.
An approximation factor of 2 for GMM was proven by induction in \cite{ravi1991facility}.
In order to prove that initializing GMM still leads to an approximation factor of 2, it is    thus sufficient to show the minimum pairwise distance in $C_{i+1}$ is larger than or equal to half of the optimal diversity of $C_i$.
Because of the separation property of the cover tree, the nodes in $C_{i+1}$ are guaranteed to have larger pairwise distance than the nodes in $C_i$. 
Using also the nesting property $C_{i+1} \subseteq C_i$, we can conclude that $div\left(C_{i+1}\right)$ is sufficiently large.
\QED

Figure~\ref{fig:theory_rl} (left) shows the approximation factor for different bases $b$. 
Especially for $b=2$, using \CTGreedy{} or \CTInherit{} as selection strategy reduces the approximation factor.
Figure \ref{fig:theory_rl} (right) shows the bound on the diversity derived in Eq. \eqref{eq:theorem1div}.
One can see, for small $\beta$ the diversity of the subset is given by the separation criterion. 
This can also be explained intuitively. As $\beta$ increases, the pairwise distance that is bounded by the separation criterion decreases. However, the cover radius $b^i$ of each node decreases as well, so that at least $k$ nodes in $C_i$ must lie close to the data points in the optimal solution to be able to cover them. Therefore, the pairwise distance of those nodes increases as $\beta$ decreases.
When the termination layer corresponds to a lower level of the tree (larger $\beta$), it is beneficial to use a GMM-based selection strategy.
The minimum of both bounds corresponds to the approximation factor.

\section{Tightness of Performance Bounds}
\label{sec:Experiment}

\begin{figure}
\centering
\begin{subfigure}{.25\textwidth}
  \centering
  \includegraphics[width=1.\linewidth]{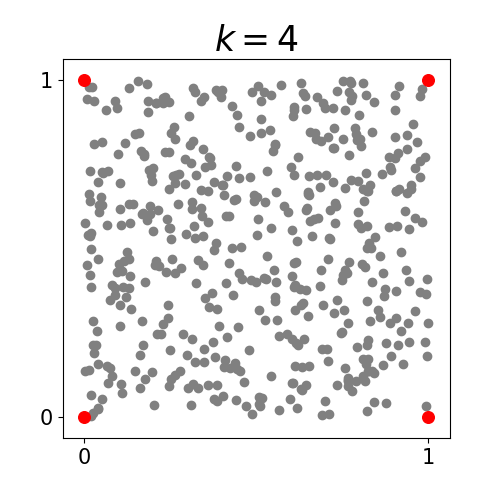}
\end{subfigure}%
\begin{subfigure}{.25\textwidth}
  \centering
  \includegraphics[width=1.\linewidth]{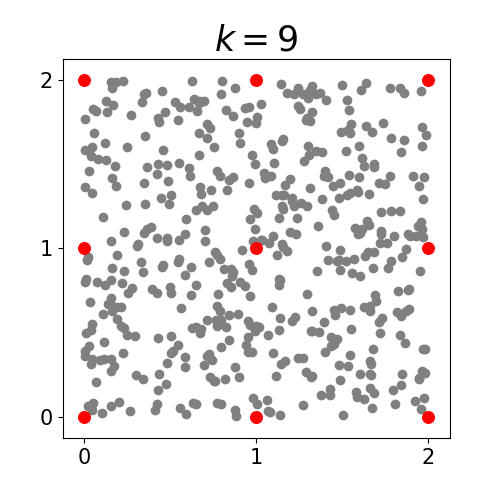}
\end{subfigure}%
\begin{subfigure}{.25\textwidth}
  \centering
  \includegraphics[width=1.\linewidth]{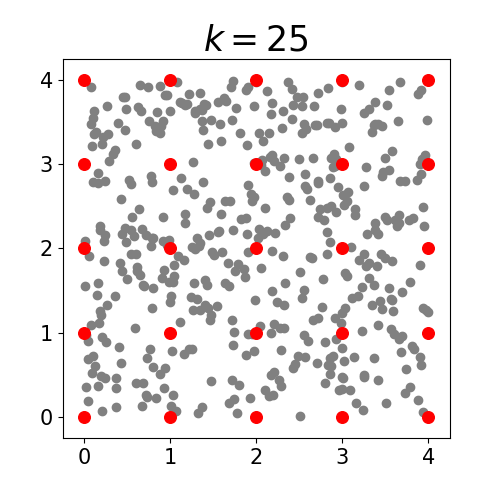}
\end{subfigure}
\caption{2D artificial data consisting of $2 \times 2$ (left), $3 \times 3$ (center) or $5 \times 5$ grid points (right) and random, uniformly distributed points.  The optimal diversity is $d^* = 1.0$.}
\label{fig:griddata}
\vspace{-3mm}
\end{figure}

According to our theory in Section~\ref{sec:Theory}, the approximation factor of ICT methods are
$\alpha^{\mathrm{\CTRandom}} = 8$,
$\alpha^{\mathrm{\CTGreedy}} = \alpha^{\mathrm{\CTInherit}} = 6$ for $b = 2$.
In this section, we investigate if these bounds are tight or whether there is still room for improvement.
% We conducted various experiments on artificial as well as on real data sets. 
% Table \ref{table:experiments} summarizes information for each experiment. 
% For the cover tree we used the implementation provided in \cite{drosou2013poikilo}.
%
%\subsection{Artificial Data}
%
We first conducted an artificial data experiment, designed for validating our theory.
We created a set of data points consisting of grid points and random points (see Fig. \ref{fig:griddata} for examples).
For $k$ equal to the number of grid points, the optimal diverse subset is given by the grid.
% \begin{table}[]
% \centering
% \caption{Summary of conducted experiments.}
% \label{table:experiments}
% \begin{tabular}{llllll}
% \toprule
% & Experiment & Dimension & Sample Size     & Diverse Set Size k & Distance Metric \\ 
% \midrule
% \multirow{2}{*}{Artificial} 
% & Grid 2D    & $2$         & $\lbrace500, 1000, 5000\rbrace$ & $\lbrace4,9,25\rbrace$             & Euclidean       \\
% & Grid 5D    & $5$         & $5000$            & $\lbrace32,243,1024\rbrace$        & Euclidean       \\ 
% \midrule
% \multirow{3}{*}{Real}
% & Cities     & $2$         & $10975$           & $[2, 100]$       & Euclidean       \\
% & Faces      & $624$       & $3840$            & $[2, 100]$        & Cosine          \\
% & MNIST      & $784$       & $70000$           & $[2, 100]$        & Cosine          \\ 
% \bottomrule
% \end{tabular}
% \end{table}
%All experiments were conducted on a Intel(R) Xeon(R) CPU E5-2640 v4 with 2.40GHz.

\begin{figure}
\centering
\begin{subfigure}{.4\textwidth}
  \centering
  \includegraphics[width=1.\linewidth]{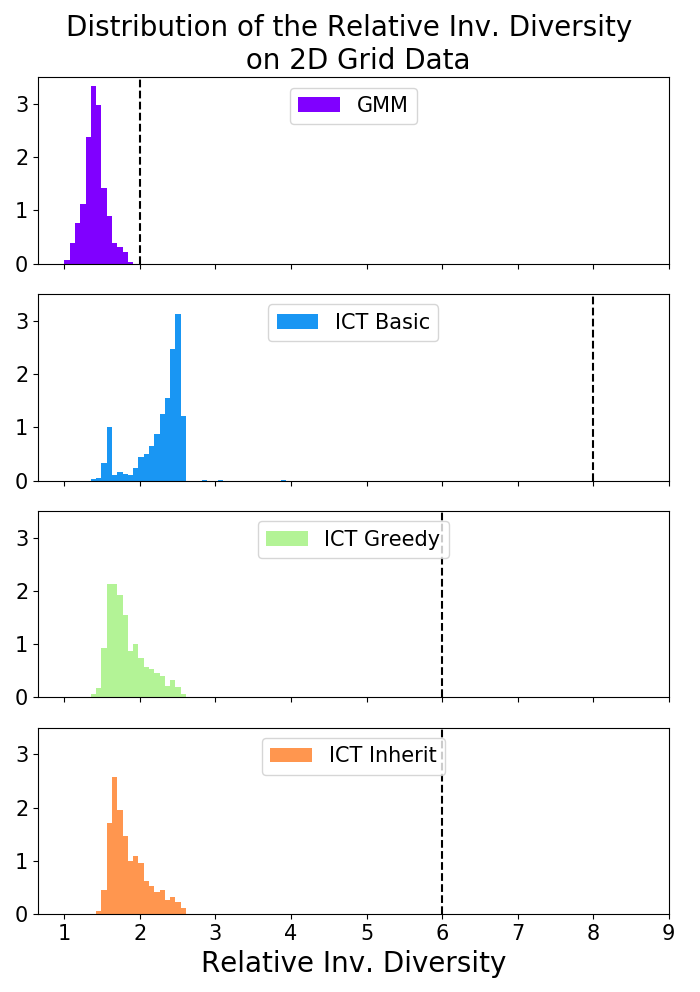}
\end{subfigure}%
\begin{subfigure}{.4\textwidth}
  \centering
  \includegraphics[width=1.\linewidth]{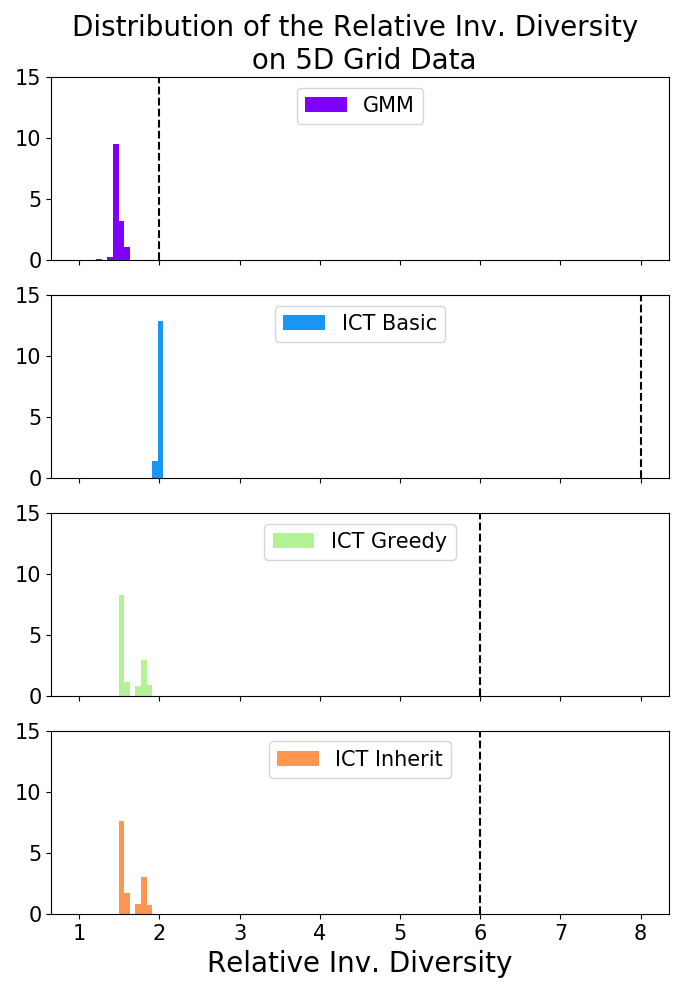}
\end{subfigure}
\caption{Estimated distribution of the \ridfull{} on the 2D (left) and 5D (right) grid data. The cover tree was built with $b=2$. The dashed vertical lines indicate the proven approximation factor.}
\label{fig:distAF}
\vspace{-3mm}
\end{figure}

Using the methods presented above, we approximate the diverse subsets and computed the \ridfull{} $\nicefrac{d^*}{S^f}$ for $f \in \lbrace \mathrm{GMM}, \mathrm{\CTRandom}, \mathrm{\CTGreedy}, \mathrm{\CTInherit} \rbrace$.
According to Eq. \eqref{eq:AF}, the computed \rid{} can at most be as large as the approximation factor and small \rid{} indicate better results. 
By repeating the experiments (100 trials each), we were able to estimate the distribution of the \rid{}. 
Figure \ref{fig:distAF} shows the result on the 2D and 5D grid data for base $b$ set to $2$. 

We still observe a big gap between the theoretical bounds and empirical observations.
Although GMM has an approximation factor that is three times lower than the guaranteed approximation factor of \CTGreedy{} and \CTInherit{} with $b=2$, only minor differences can be seen.
One can see, that the center of the distribution of the observed \rid{} is even lower than 2.
The highest observed \rid{} was $2.9$ for \CTInherit, $2.6$ for \CTGreedy{} and $3.9$ for \CTRandom{} on the 2D data set.
Note that for larger $k$, we tend to select nodes from lower layers of the tree.
Lower layers hold more, but evenly spaced nodes, because their cover radius $b^i$ is smaller.
This might explain the excellent performance on the 5D grid data.

The approximation factor is defined as an upper bound. 
Thus, when an approximation factor is proven, it does not imply, that no smaller approximation factor is possible.
Because of the high performance of the cover tree approaches on the grid data sets, one might expect lower approximation factors, than the ones that were proven in Sec. \ref{sec:Theory}.
However, we provide two examples, that show the tightness of $\alpha^\mathrm{\CTRandom}$, $\alpha^\mathrm{\CTGreedy}$ and $\alpha^\mathrm{\CTInherit}$ (see Appendix \ref{sec:tightness}).
Thereby, we have shown that no lower approximation factor can be proven.
%The example in the Appendix are specifically designed so that we observe the worst possible \rid{}.
%Recall, that the structure of the cover tree depends on the order in which the data is incrementally inserted. 
Note, that the \rid{} does not only depend on the data pool, but also on the order in which the data is added to the tree.
Even for the examples, which are discussed in the Appendix, we only get the worst case \rid{}, if the data points are added in a specific order.
When this order is changed, we might also get the optimal diverse set.
See the Appedendix \ref{sec:insertion} for the incremental insertion algorithm of the cover tree.
With the excellent performance on the artificial data and this observation at hand, we can conclude, that observing the worst possible \rid{} is highly unlikely, but not impossible.

%\subsection{Real Data}
%\label{sec:Experiment.RealData}
We also conducted {\bf real world data experiments} - both high dimensional and large sample sized.
\cities{} \cite{cities} consists of the latitude and longitude of cities in Europe. 
\faces{} \cite{faces} consists of 60x64 grey-scale images of people taken with varying pose and expressions.
\mnist{} \cite{mnist} holds 28x28 images of handwritten digits.
For each of the data sets we ran GMM and the cover tree approaches (built with $b=2$) with varying diverse set sizes ($k \in [2, 100]$).
In general the optimal diversity is unknown.
Therefore, we plotted the inverse diversity relative to the diversity of GMM, i.e.,
%\begin{equation}
%\textstyle
$\nicefrac{div(S^{GMM})}{div(S^f)}$,
%\end{equation}
where $f \in \lbrace \mathrm{\CTRandom}, \mathrm{\CTGreedy}, \mathrm{\CTInherit} \rbrace$ corresponds to the applied method.
This can be used to bound the true \rid.
If, for example, $\nicefrac{div(S^{GMM})}{div(S^{\mathrm{\CTInherit}})} = 1.5$, we can conclude $\nicefrac{d^*}{div(S^{\mathrm{\CTInherit}})} \leq 3$, because GMM has an approximation factor of 2.

Figure \ref{fig:Af_Ct_first} shows the inv. diversity relative to the diversity of GMM on the \cities, \faces{} and \mnist{} data set.
%Once again, 
%we can see the high performance of the cover tree approaches. 
%Both \CTGreedy{} and \CTInherit{} give \rid{} wrt to GMM that range mostly between 1 and 1.1, meaning that both methods and GMM compute subsets with diversities of almost the same magnitude.
\CTGreedy{}, \CTInherit{} and GMM compute subsets with diversities of almost the same magnitude.
Even \CTRandom{} gives results with high diversity. 
%Its worst \rid{} is only marginally larger than 2, so that $\nicefrac{d^*}{div(S^{\mathrm{\CTRandom}})} < 4.2$.
%Compared to \CTGreedy{} or \CTInherit, \CTRandom{} gives more unstable results.
Note, that the inv. diversity relative to GMM of \CTRandom{} resembles a step-function.
This can be explained by the layer-wise structure of the cover tree. 
When $k$ is small compared to the number of nodes in the termination layer, \CTRandom{} might select poorly which results in lower diversity than \CTInherit.
As $k$ approaches the size of the layer, both approaches will select similar subsets. 

% \begin{figure}
%   \centering
%   \includegraphics[width=0.95\textwidth]{Real.png}
%   \caption{Computation time in ms on the \cities, \faces{} and \mnist data set.}
% \label{fig:Af_Ct}
% \end{figure}
% Figure \ref{fig:Af_Ct} shows the computation time of selecting the diverse subsets for the different approaches.
% One can clearly see the efficiency of the cover-tree approaches.
% Compared to GMM, \CTRandom, \CTGreedy{} and \CTInherit{} have fast computation time even for large $k$.
% The computation time of the cover tree approaches shows a step function behavior.
% Again, this can be explained by the layer-wise structure of the cover tree. 
% When the termination layer holds exactly $k$ nodes, no selection must be made. 
% We can also see the difference in the complexity of \CTGreedy{} and \CTInherit. 
% When the layer before the termination layer holds almost $k$ nodes, i.e. $\vert C_{i+1} \vert \approx \vert C_i \vert$, \CTInherit{} only has to select few nodes. 

\section{Application to Image Generator}

\begin{figure}[t]
\centering
\begin{subfigure}{0.45\textwidth}
\centering
\includegraphics[width=0.9\textwidth]{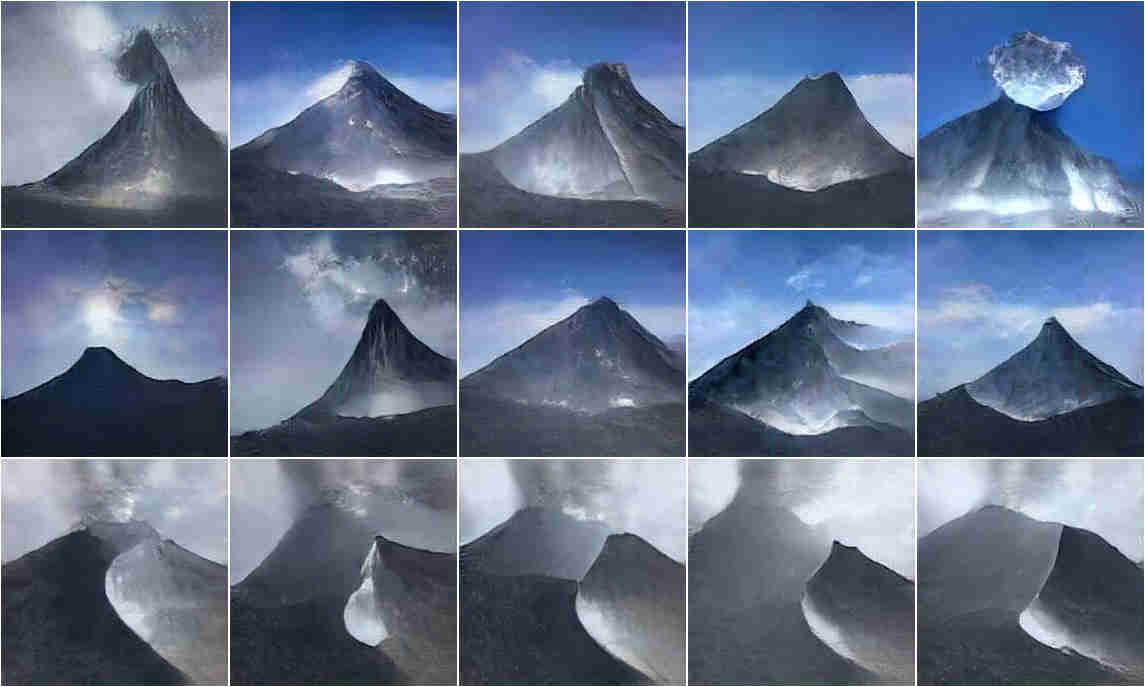}
\end{subfigure}
\centering
\begin{subfigure}{0.45\textwidth}
\includegraphics[width=0.9\textwidth]{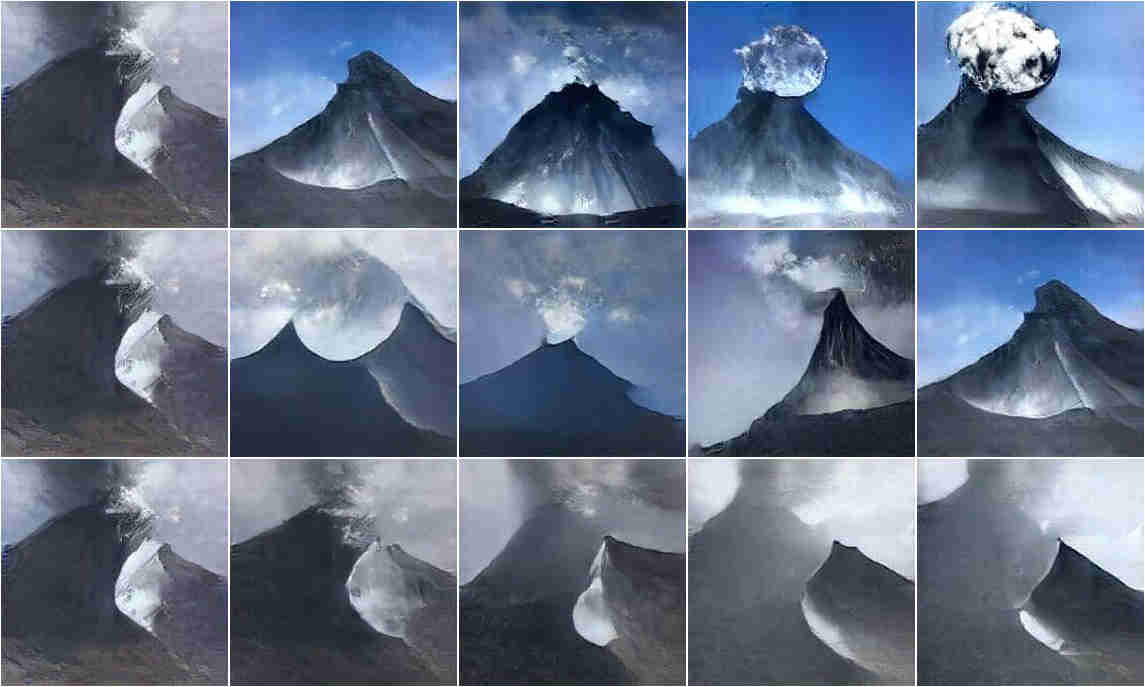}
\end{subfigure}
\caption{$k=5$ randomly chosen images (left) from the MCMC sequence of PPGN
for \emph{Volcano} class, and the corresponding  diverse sets (right) chosen by \CTInherit,
after 20 (bottom), 100 (middle), and 200 (top) MCMC steps.}
\label{fig:vulcano}
\vspace{-3mm}
\end{figure}

As a new use of dynamic diversification, we applied \CTInherit{}
to sequentially collect a diverse set from the MCMC image sequence generated by the \emph{plug and play generative networks} (PPGN) \cite{PPGN2017},%
\footnote{
The code is available from \url{https://github.com/Evolving-AI-Lab/ppgn}.
}
which have shown to provide 
highly diverse realistic images with high resolution.
For generated images by MCMC sampling, 
there is strong correlation between subsequent images.
One needs to cherry-pick diverse images by hand,
or generate a long sequence and randomly choose images from it,
to show how diverse the images are, that a new sampling strategy generates.
Dynamic diversification by \CTInherit{} can automate this process,
and might be used as a standard tool to assess the diversity of samplers.

Fig.\ref{fig:vulcano} shows images of the \emph{Volcano} class generated by PPGN.
We ran PPGN up to 200 steps, adding a new image to \CTInherit{} in each step.
In the right half of Fig.\ref{fig:train}, we show the $k=5$ diverse images
after 20 (bottom row), 100 (middle row), and 200 (top row) steps of image generation.
For comparison, we show randomly chosen images in the left half.
We can see that \CTInherit{} successfully chose more diverse images than the random choice.
More examples for other classes are shown in Appendix~\ref{sec:MoreExamplesPPGN}.

\section{Conclusion}

Selecting a diverse subset of a dynamic sample pool---dynamic diversification---is a common problem in many machine learning applications.
The diversification problem is known to be NP-hard, 
but polynomial time approximation algorithms, such as \GMMfull (GMM) algorithm,
can have an approximation factor of 2, i.e., they are guaranteed to give a solution with diversity larger than half of the optimal diversity.
However, GMM has to be performed from scratch every time the sample pool is updated, e.g., new items are added or old items are removed,
and therefore is not suitable for dynamic diversification.
Previous work argued that 
cover trees, originally proposed for nearest neighbor search, 
could be adapted for dynamic diversification,
and proposed 
\CTRandom, \CTGreedy{} and \CTInherit{},
which gave results only marginally worse than the results of GMM, while the approximation factor of those approaches was assumed to be four times larger.

In this work we have conducted both theoretical analyses and extensive experiments to fill the gap between the theoretical bound and empirical observation.
Specifically, we could prove a tighter approximation factor for \CTGreedy{} and \CTInherit, reducing it to 6 instead of 8 for a cover tree with base $b=2$.
Through artificial experiment, we have validated the bounds, and assessed the tightness of the bounds.
Even on real world data sets, all three cover tree approaches give excellent results, with diversities almost of the same magnitude as the diversity given by GMM.
The performance of the cover tree approaches is remarkably higher than the theoretical approximation factor, which might imply that our bounds are still loose.
However, we found worst case examples that achieve the theoretical approximation factor, which proves that our bounds for ICT are \emph{tightest possible}.

In general, the diversity of subsets computed with one of the cover tree approaches does not only depend on the data pool, but also on the order in which the data is added to the tree. 
Therefore, we conclude, that observing the worst possible \ridfull{} is highly unlikely.
Further effort must be made to assess how likely cover tree approaches give solutions with worst possible diversity.

Finally, our demonstration of dynamic diversification for generative image samplers shows the high potential of our theoretical insight for practical applications. Future work will also study diversification in scientific applications, where systematically creating a diverse (and therefore representative) subset of large data corpora can lead to novel insight, e.g. in molecular dynamics or sampling applications in chemistry.

\subsubsection*{Acknowledgments}
%This work was supported by
%the German Research Foundation (GRK 1589/1)
%by the Federal Ministry of Education and Research (BMBF) 
%under the project Berlin Big Data Center (FKZ 01IS14013A).
%This work was supported by the Fraunhofer Society under the MPI-FhG collaboration project (600393).
The authors thank Vignesh Srinivasan of Fraunhofer HHI for discussion on the image generator application. 
This work was supported by the German Research Foundation (GRK
1589/1) by the Federal Ministry of Education and Research (BMBF) under the project Berlin Big Data Center (FKZ 01IS14013A).

\small
\bibliography{MachineLearning,MachineLearning2,Hannah}

\begin{thebibliography}{10}

\bibitem{abbar2013diverse}
Sofiane Abbar, Sihem Amer-Yahia, Piotr Indyk, Sepideh Mahabadi, and Kasturi~R
  Varadarajan.
\newblock Diverse near neighbor problem.
\newblock In {\em Proceedings of the twenty-ninth annual symposium on
  Computational geometry}, pages 207--214. ACM, 2013.

\bibitem{faces}
The UCI~KDD Archive.
\newblock {CMU} faces images.
\newblock \url{http://kdd.ics.uci.edu/databases/faces/faces.html}.
\newblock Accessed on 2018-05-18.

\bibitem{Beygelzimer06}
Alina Beygelzimer, Sham Kakade, and John Langford.
\newblock Cover trees for nearest neighbor.
\newblock In {\em Proc. of ICML}, pages 97--104, 2006.

\bibitem{beygelzimer2006cover}
Alina Beygelzimer, Sham Kakade, and John Langford.
\newblock Cover trees for nearest neighbor.
\newblock In {\em Proceedings of the 23rd international conference on Machine
  learning}, pages 97--104. ACM, 2006.

\bibitem{cities}
{Center For International Earth Science Information Network-CIESIN-Columbia
  University}.
\newblock Gridded population of the world: Basic demographic characteristics.
\newblock \url{https://doi.org/10.7927/H45H7D7F}.
\newblock Accessed on 2016-12-13.

\bibitem{Demidova10}
E.~Demidova, P.~Fankhauser, X.~Zhou, and W.~Nejdl.
\newblock Div{Q}: {D}iversification for keyword search over structured
  databases.
\newblock In {\em Proc. of SIGIR}, pages 331--338, 2010.

\bibitem{drosou2013poikilo}
Marina Drosou and Evaggelia Pitoura.
\newblock Poikilo: A tool for evaluating the results of diversification models
  and algorithms.
\newblock {\em Proceedings of the VLDB Endowment}, 6(12):1246--1249, 2013.

\bibitem{drosou2014diverse}
Marina Drosou and Evaggelia Pitoura.
\newblock Diverse set selection over dynamic data.
\newblock {\em IEEE Transactions on Knowledge and Data Engineering},
  26(5):1102--1116, 2014.

\bibitem{Erkut90}
E.~Erkut.
\newblock The discrete $p$-dispersion problem.
\newblock {\em European Journal of Operational Research}, 1(1), 1990.

\bibitem{Erkut94}
E.~Erkut, Y.~Ulkusal, and O~Yenicerioglu.
\newblock A comparison of $p$-dispersion heuristics.
\newblock {\em Computers and Operations Research}, 21(10), 1995.

\bibitem{Gollapudi09}
S.~Gollapudi and A.~Sharma.
\newblock An axiomatic approach for result diversification.
\newblock In {\em Proc. of WWW}, pages 381--390, 2009.

\bibitem{Izbicki15}
M.~Izbicki and C.~R. Shelton.
\newblock Fast cover trees.
\newblock In {\em Proc. of ICML}, 2015.

\bibitem{kunaver2017diversity}
Matev{\v{z}} Kunaver and Toma{\v{z}} Po{\v{z}}rl.
\newblock Diversity in recommender systems--a survey.
\newblock {\em Knowledge-Based Systems}, 123:154--162, 2017.

\bibitem{mnist}
Yann LeCun and Corinna Cortes.
\newblock {MNIST} handwritten digit database.
\newblock http://yann.lecun.com/exdb/mnist/, 2010.

\bibitem{Liu09}
Z.~Liu, P.~Sun, and Y.~Chen.
\newblock Structured search result differentiation.
\newblock In {\em Proc. of VLDB}, 2009.

\bibitem{Minack11}
E.~Minack, W.~Siberski, and W.~Nejdl.
\newblock Incremental diversification for very large sets: {A} streaming-based
  approach.
\newblock In {\em Proc. of SIGIR}, 2011.

\bibitem{PPGN2017}
A.~Nguyen, J.~Clune, Y.~Bengio, A.~Dosovitskiy, and J.~Yosinski.
\newblock Plug \& play generative networks: Conditional iterative generation of
  images in latent space.
\newblock In {\em 2017 IEEE Conference on Computer Vision and Pattern
  Recognition (CVPR)}, pages 3510--3520, 2017.

\bibitem{ravi1991facility}
SS~Ravi, Daniel~J Rosenkrantz, and Giri~Kumar Tayi.
\newblock Facility dispersion problems: Heuristics and special cases.
\newblock In {\em Workshop on Algorithms and Data Structures}, pages 355--366.
  Springer, 1991.

\bibitem{Vee08}
E.~Vee, U.~Srivastava, J.~Shanmugasundaram, P.~Bhat, and S.~A. Yahia.
\newblock Efficient computation of diverse query results.
\newblock In {\em Proc. of ICDE}, pages 228--236, 2008.

\bibitem{Vieira11}
M.~R. Vieira, H.~L. Razente, M.~C.~N. Barioni, M.~Hadjieleftheriou,
  D.~Srivastava, C.~T. Jr., and V.~J. Tsotras.
\newblock On query result diversification.
\newblock In {\em Proc. of ICDE}, 2011.

\bibitem{williamson2011design}
David~P Williamson and David~B Shmoys.
\newblock {\em The design of approximation algorithms}.
\newblock Cambridge university press, 2011.

\bibitem{Yu09}
C.~Yu, L.~V. Lakshmanan, and S.~Amer-Yahia.
\newblock Recommendation diversification using explanations.
\newblock In {\em Proc. of ICDE}, 2009.

\bibitem{Zhang08}
M.~Zhang and N.~Hurley.
\newblock Avoiding monotony: {I}mproving the diversity of recommendation lists.
\newblock In {\em Proc. of RecSys}, 2008.

\bibitem{Ziegler05}
C.-N. Ziegler, S.~M. McNee, J.~A. Konstan, and G.~Lausen.
\newblock Improving recommendation lists through topic diversification.
\newblock In {\em Proc. of WWW}, pages 381--390, 2005.

\end{thebibliography}
\bibliographystyle{plain}
\normalsize

\newpage

\appendix

\section{Incremental Insertion}\label{sec:insertion}
Algorithm \ref{alg:insert} shows how a data point $p$ can be inserted in an existing cover tree. 
It is based on the insertion algorithm in \cite{beygelzimer2006cover}, which was extended to arbitrary bases in \cite{drosou2014diverse}.
We reduced the complexity of the algorithm by omitting the nearest-parent heuristic.
\begin{algorithm}[H]
\begin{algorithmic} [1]
\REQUIRE $p$ - data point $\; \; \; Q_i$ - subset of nodes of layer $i$ of the cover tree
\ENSURE Cover tree with $p$ inserted

\STATE $C \leftarrow \left\lbrace children(q) \vert q \in Q_i \right\rbrace$

\IF{$\min_{c \in C} d(p,c) > b^i$}
	\RETURN \TRUE
\ELSE
	\STATE $Q_{i-1} \leftarrow \left\lbrace c \in C \vert d(p,c) \leq 			\frac{b^i}{b-1} \right\rbrace$
	\STATE flag $\leftarrow$ Insert$(p, Q_{i-1}, i-1)$
	\IF{flag \AND $d(p, Q_i) \leq b^i$}
		\STATE $q^* \leftarrow$ choose node in $Q_i$ as parent
		\STATE make $p$ a child of $q^*$
		\RETURN \FALSE
	\ELSE
		\RETURN flag
	\ENDIF
\ENDIF
\end{algorithmic}
\caption{Insert($p$, $Q_i$, $i$)}
\label{alg:insert}
\end{algorithm}
We start with the root layer Insert$(p, C_{i_{max}}, i_{max})$ and traverse the cover tree until we find the first layer such that $p$ is separated from all other nodes (line 2).
Only the nodes that can cover $p$ are considered in each iteration ($Q_i$).
When we found an appropriate layer (line 7), we add $p$ as a child to one  node in $Q_i$ (line 8,9).

\section{Proof of Theorem 1}\label{sec:prooftheorem1}
\subsection{Bound on $\beta$}
Recall (Eq. \eqref{eq:beta}), that $\beta$ is defined as the fraction of the optimal diversity and the cover radius of the termination layer. Because the optimal solution is not known, the exact value of $\beta$ can in general not be identified. We will use the properties of the cover tree to give a bound on possible values of $\beta$.
We first consider how $\beta$ must be chosen, so that $C_i$ holds at least $k$ nodes. To do so, assume we try to enclose, two data points $p$ and $p'$ with $d(p, p') = d^*$ with one ball. A single ball won't be able to enclose those two data points, if its radius is less than $\nicefrac{d^*}{2}$. No matter where the ball is placed, the maximal distance it can enclose is \textit{less} than $2 \cdot \nicefrac{d^*}{2}$ and thus, smaller than the distance between $p$ and $p'$.
Because it holds for any pair of data points, the same will hold for the general case of $k$ data points. If the pairwise distance between two data points is at least $d^*$, any ball with radius less than $\nicefrac{d^*}{2}$ can only enclose one of those $k$ data points at once. Thus, we need at least $k$ balls to enclose $k$ data points. 

Let $desc(c)$ denote the the set of descendants of node $c$. 
We say, that a node $c$ \emph{covers} a node $v$, if $v \in desc(c)$.
Furthermore, we say that a layer $C$ of the cover tree \emph{covers} a set of nodes $C'$, if
\begin{equation}
C' \subseteq \left( \bigcup_{c \in C} desc(c) \right).
\end{equation}
As a consequence of the covering and the nesting property, each layer of the cover tree does not only cover the layer beneath it, but covers the whole data set $V$, i.e.
\begin{equation}
\forall i \in [i_{\min},  i_{\max}] \, : \left( \bigcup_{c \in C_i} desc(c) \right) = V.
\end{equation}
Let $S^* \subseteq V$ with $div(S^*) = d^*$ denote the optimal diverse set. Because $S^*$ is a subset of $V$, $C_i$ will cover the optimal solution i.e. 
\begin{equation}
S^* \subseteq \bigcup_{c \in C_i} desc(c).
\end{equation}
The maximal distance to any descendant of any node $c \in C_i$ can be bounded
\begin{equation}
\forall c \in C_i, \forall v \in desc(c): d(c,v) < \frac{b^{i+1}}{b-1} =  \frac{d^*}{\beta} \frac{b}{b-1}.
\end{equation}
Thus, if the maximal distance to any descendant is less than $\nicefrac{d^*}{2}$, any node $c$ in $C_i$ will only be able to cover one of the optimal data points in $S^*$. Because $|S^*| = k$, $C_i$ is guaranteed to hold at least $k$ nodes, if
\begin{align}
 \frac{b^{i+1}}{b-1} &\leq \frac{d^*}{2} \\
\Leftrightarrow  \frac{d^*}{\beta} \frac{b}{b-1} &\leq \frac{d^*}{2} \\
\Leftrightarrow  \frac{2b}{b-1} &\leq \beta
\end{align}
(we can use $\leq$ instead of strict inequality, because $\nicefrac{b^{i+1}}{b-1}$ is already a strict upper bound for the distance to the descendants). 

We have shown, that $C_i$ will hold at least $k$ data points, if $b^i \leq d^* \cdot \nicefrac{b-1}{2b}$. 
Note, that the cover tree might already hold $k$ nodes in an higher layer, especially because $C_i$ must not only cover $S^*$ but the whole data set $V$. In those cases, the diversity of the solution will be higher (separation criterion).

This property can be used to find an upper bound for $\beta$. Assume $C_{i}$ is the termination layer and $b^{i} = d^* \cdot \nicefrac{b-1}{2b^2}$. In that case 
$b^{i+1} = b \cdot d^* \cdot \nicefrac{b-1}{2b^2} = d^* \cdot\nicefrac{b-1}{2b}$.
However, as shown above, this implies that $C_{i+1}$ holds at least $k$ nodes. Thus, $C_i$ cannot be the termination layer, because $C_{i+1}$ would have been. We can conclude
\begin{equation}
\beta < \frac{2b^2}{b-1}.
\end{equation}

\subsection{Bound on the Diversity}
We prove each part of the maximum in Eq. \eqref{eq:theorem1div} individually. The first part follows immediately from the separation criterion and the definition of $\beta$. Any subset of $C_i$ will have a diversity of at least $\nicefrac{d^*}{\beta}$,
\begin{equation}
\forall c, c' \in C_i, d(c, c') > b^i \Rightarrow div(C_i) > \frac{d^*}{\beta}.
\end{equation}
In order to prove the second part of the maximum, we assume every data point of the optimal solution is covered by a single node in $C_i$, i.e. $\beta \geq \nicefrac{2b}{b-1}$.
Let $c,c' \in C_i$ be an arbitrary pair of nodes in the termination layer which covers $p,p' \in S^*$. Without loss of generality we assume $c$ covers $p$, $c'$ covers $p'$ and $d(p,p') \geq d^*$. Because of the covering property 
\begin{align}
d(c,p) &< \frac{b^{i+1}}{b-1} = \frac{d^*}{\beta} \frac{b}{b-1}\\
d(c',p') &< \frac{b^{i+1}}{b-1} = \frac{d^*}{\beta} \frac{b}{b-1}.
\end{align}
Using the triangle inequality, we get
\begin{align}
d(p,c) + d(c,p') &\geq d(p,p') \\
d(p,c) + d(c,c') + d(c',p') &\geq d(p,p')\\
d(c,c') &\geq d(p,p') - d(p,c) - d(c',p') \\
d(c,c') &> d^* - \frac{d^*}{\beta} \frac{b}{b-1} - \frac{d^*}{\beta} \frac{b}{b-1} \\
d(c,c') &> d^* (1 - \frac{1}{\beta}\frac{2b}{b-1}).
\end{align}
Because any data point of the optimal solution is by assumption covered by one node in $C_i$ and the pairwise distance of those data points is at least $d^*$, it follows that $C_i$ holds a subset of $k$ nodes, that have pairwise distance of at least $d^* (1 - \nicefrac{1}{\beta} \cdot \nicefrac{2b}{b-1})$. 

Recall, that we assumed $\beta \geq \nicefrac{2b}{b-1}$ in the beginning of the second part of the proof, to make sure that every data point of the optimal solution is covered by a different node of the termination layer. This does not impose any restrictions on the validity of the bound, because for $\beta < \nicefrac{2b}{b-1}$, $\text{max} \left\lbrace  \nicefrac{d^*}{\beta}, \, d^* (1 - \nicefrac{1}{\beta}\cdot\nicefrac{2b}{b-1})  \right\rbrace = \nicefrac{d^*}{\beta}$ and the overall bound still holds.
This concludes the proof of the theorem.

\subsection{Fraction of Diversities}\label{sec:CTfractionofdiversities}
First note, that $\nicefrac{d^*}{\beta}$ decreases and $d^* \cdot (1 - \nicefrac{1}{\beta} \cdot \nicefrac{2b}{b-1})$ increases as $\beta$ increases.
As a consequence, the bound of the diversity in Eq. \eqref{eq:theorem1div} is minimal, when
\begin{equation}
\frac{d^*}{\beta} = d^* (1 - \frac{1}{\beta}\frac{2b}{b-1}).
\end{equation}
Solving for $\beta$ leads to
\begin{align}
\frac{1}{\beta} &= 1 - \frac{1}{\beta}\frac{2b}{b-1}\\
\frac{1}{\beta} (1 + \frac{2b}{b-1}) &= 1\\
1 + \frac{2b}{b-1} &= \beta.
\end{align}
We use this minimum and give a bound on the fraction of the diversities
\begin{align}
\frac{d^*}{div(C)} &\leq \frac{d^*}{\min_{\beta} \left\lbrace \max \left\lbrace  \frac{d^*}{\beta}, d^* (1 - \frac{1}{\beta}\frac{2b}{b-1})\right\rbrace \right\rbrace}\\
&= \frac{d^*}{\frac{d^*}{1 + \frac{2b}{b-1}}}\\
&= 1 + \frac{2b}{b-1}.
\end{align}

\section{Proof of Corollary of Theorem 1}\label{sec:proofcorollarly1}
Suppose we use \CTGreedy{} to select a subset of $k$ nodes from $C_i$. The diversity of the selected subset cannot be worse than $\nicefrac{d^*}{\beta}$, because that bound is given by the separation criterion. As it was shown in \cite{ravi1991facility}, GMM has an approximation factor of $2$. We select $k$ nodes from $C_i$ not from $V$ itself. Thus, the best possible diversity is not $d^*$, but rather given by Eq. \eqref{eq:theorem1div}, i.e. the optimal diversity in layer $C_i$. We get what is stated in Eq. \eqref{eq:ctgreedydiv}.

As it was shown in \cite{ravi1991facility}, the complexity of GMM is in $\mathcal{O}(N \cdot k)$ where $N$ is the size of the set, from which $k$ data points are selected. Here we select data points from the termination layer, i.e. $N = \vert C_i\vert$.
Analogous to what was shown in Sec. \ref{sec:CTfractionofdiversities}, we can prove the approximation factor of \CTGreedy. Solving
\begin{equation}
\frac{d^*}{\beta} = \frac{1}{2} \, d^* (1 - \frac{1}{\beta}\frac{2b}{b-1})
\end{equation}
for $\beta$ leads to
\begin{equation}
\beta = 2 + \frac{2b}{b-1}.
\end{equation}
Again, we use this value for the calculation of the minimum of $div(S^{\mathrm{\CTGreedy}})$
\begin{align}
\frac{d^*}{div(S^{\mathrm{\CTGreedy}})} &\leq \frac{d^*}{\min_{\beta} \left\lbrace \max \left\lbrace  \frac{d^*}{\beta}, \frac{1}{2} d^* (1 - \frac{1}{\beta}\frac{2b}{b-1})\right\rbrace \right\rbrace}\\
&= \frac{d^*}{\frac{d^*}{2 + \frac{2b}{b-1}}}\\
&= 2 + \frac{2b}{b-1}.
\end{align}

\section{Proof of Theorem 2}\label{sec:prooftheorem2}
\CTInherit{} uses GMM to calculate a subset of layer $C_i$.
Instead of starting with a randomly selected data point (as in GMM or \CTGreedy{} resp.), we start with $C_{i+1}$ and subsequently add new data points.
The proof of the approximation factor of GMM relies on induction, thereby it was shown that the approximation factor holds after each iteration (see \cite{ravi1991facility} for more details). 
Once it is shown that $C_{i+1}$ does not contradict an approximation factor of 2, the remaining proof in \cite{ravi1991facility} can simply be applied.
Thus, we show for any possible value of $\beta$ the minimum possible pairwise distance in $C_{i+1}$ is larger than or equal to half of the optimal diversity in $C_i$. 
Because the diversity of $C_i$ can at most be as large as the optimal diversity, $\beta \geq 1$ (recall Eq. \eqref{eq:CTboundsep}).
As it was stated in the first theorem, $\beta < \nicefrac{2b^2}{b-1}$.
According to the separation criterion
\begin{equation}
\forall c, c' \in C_{i+1}, d(c, c') > b^{i+1} = b \cdot \frac{d^*}{\beta}
\end{equation} and thus, we show
\begin{equation}
\forall \beta \in \left[1, \frac{2b^2}{b-1} \right[ \; . \; b \cdot \frac{d^*}{\beta} \geq \text{max} \left\lbrace  \frac{d^*}{\beta}, \frac{1}{2} \, d^* (1 - \frac{1}{\beta}\frac{2b}{b-1})  \right\rbrace.
\end{equation}
\begin{itemize}
\item[Case A] $\text{max} \left\lbrace  \frac{d^*}{\beta}, \frac{1}{2} \, d^* (1 - \frac{1}{\beta}\frac{2b}{b-1})  \right\rbrace = \frac{d^*}{\beta}$ 
\begin{equation}
b \cdot \frac{d^*}{\beta} \geq \frac{d^*}{\beta}
\end{equation} 
holds, since $b > 1$.
\item[Case B] $\text{max} \left\lbrace  \frac{d^*}{\beta}, \frac{1}{2} \, d^* (1 - \frac{1}{\beta}\frac{2b}{b-1})  \right\rbrace = \frac{1}{2} \, d^* (1 - \frac{1}{\beta}\frac{2b}{b-1}) $ 
\begin{align}
b \cdot \frac{d^*}{\beta} &\geq  \frac{1}{2} \, d^* (1 - \frac{1}{\beta}\frac{2b}{b-1})\\
b \cdot \frac{d^*}{\beta} &\geq  \frac{1}{2} d^* - \frac{d^*}{\beta}\frac{b}{b-1}\\
b \cdot \frac{1}{\beta} &\geq  \frac{1}{2} - \frac{1}{\beta}\frac{b}{b-1}\\
b + \frac{2b}{b-1} &\geq \frac{\beta}{2}\\
2b + \frac{4b}{b-1} &\geq \beta\\
\frac{2b\cdot(b+1)}{b-1} &\geq \beta\\
\end{align} 
holds for $\beta \in \left[ 1, \frac{2b^2}{b-1} \right[$. 
\end{itemize}
Thus, 
\begin{equation} div(C_{i+1}) \geq \text{max} \left\lbrace  \frac{d^*}{\beta}, \frac{1}{2} \, d^* (1 - \frac{1}{\beta}\frac{2b}{b-1})  \right\rbrace\end{equation} which was to be shown.

When GMM is initialized with $C_{i+1}$ it only has to select $k - \vert C_{i+1}\vert$ data points. It is easy to see that the complexity reduces. In the first iteration, the distances from each data point in $C_{i+1}$ to every data point in $C_i \setminus C_{i+1}$ is computed in order to find the next data point. In each subsequent iteration, only the distance between the data point which was selected in the last iteration and the remaining available data set is computed. In total, \CTInherit{} performs
 \begin{equation} \left(\vert C_i \setminus C_{i+1}\vert \cdot (k - \vert C_{i+1}\vert)\right) + \sum_{m = 1}^{k-\vert C_{i+1}\vert-1} \left( \vert C_i \setminus C_{i+1}\vert - m\right) \end{equation}
distance computations and has therefore a complexity of $\mathcal{O}\left(\vert C_i\setminus C_{i+1}\vert\cdot \left(k - \vert C_{i+1}\vert\right)\right)$.

Because \CTInherit{} has the same bound on the diversity as \CTGreedy{} (cf. \eqref{eq:ctgreedydiv} and \eqref{eq:ctinherit_div}), both approaches have the same approximation factor.

\section{Tightness of the Proven Approximation Factors}\label{sec:tightness}
The approximation factor is defined as an upper bound.
Thus, when an approximation factor is proven, it does not imply, that no smaller approximation factor is possible. In this section we give two examples, that show the tightness of the derived approximation factors for arbitrary bases and $k=2$.
\subsection{Tightness of $\alpha^{\mathrm{\CTRandom}}$}\label{sec:thightnessRandom}
Let the data set be defined as
\begin{equation}
V = \left\lbrace  \left(\begin{array}{c} v\\ 0\\ \end{array} \right), \left(\begin{array}{c} 0\\ b + \mu\\ \end{array} \right) \vert v \in \pm \left( \lbrace 0 \rbrace \; \cup \; \left(\bigcup_{J \in [1, J_{max}]} \left\lbrace \sum_{j=1}^{J} b^2 \cdot {\left(\frac{1}{b}\right)}^{j-1} \right\rbrace \right) \; \cup \; \left\lbrace  \frac{b^3}{b-1}-\eta \right\rbrace \right) \right\rbrace,  
\end{equation}
where $\mu, \eta > 0$ small and $J_{max}$ be the smallest integer such that
\begin{equation}\label{eq:eta}
\frac{b^{3-J_{max}}}{b-1} - b^{i_{min}+1} \leq \eta.
\end{equation}
Most of the data points in $V$ lie on the line between ${(-\frac{b^3}{b-1} + \eta, 0)}^{T}$ and ${(\frac{b^3}{b-1} - \eta, 0)}^{T}$, where the distance between neighboring data points decreases as we move to the ends of the line.
Only the data point ${(0, b + \mu)}^{T}$ does not lie on that line, but above the midpoint ${(0,0)}^T$.
We choose the Euclidean distance as distance metric.
The optimal solution $S^* \subseteq V$ for $k=2$ is given by 
\begin{equation}
S^* = \lbrace {(-\frac{b^3}{b-1} + \eta, 0)}^{T}, {(\frac{b^3}{b-1} - \eta, 0)}^{T}\rbrace
\end{equation}
with a diversity of
\begin{equation}
d^* = 2 \frac{b^3}{b-1} - 2 \eta \;\; \approx \;\; 2 \frac{b^3}{b-1}.
\end{equation}
The cover tree for $V$ is built with base $b$ such that data points are incrementally added, ordered according to their absolute $x$ value.
As a result, we get the following cover tree:
\begin{itemize}
\item $i_{max}$: $C_{i_{max}} = \left\lbrace {\left( 0, 0\right)}^{T} \right\rbrace$
\item $i = 1$: $C_i = C_{i+1} \cup \left\lbrace {(0, b + \mu)}^{T}, -b^2, b^2 \right\rbrace$
\item $1>i>i_{min}$: $C_i = C_{i+1} \cup \left\lbrace {\left(-\sum_{j=1}^{i_{max} - i} b^2 \cdot {\left(\frac{1}{b}\right)}^{j-1},0\right)}^T,  {\left(\sum_{j=1}^{i_{max} - i} b^2 \cdot {\left(\frac{1}{b}\right)}^{j-1},0\right)}^T \right\rbrace$
\item $i_{min}$: $C_{i_{min}} = C_{i_{min} + 1} \cup \left\lbrace {(-\frac{b^3}{b-1} + \eta, 0)}^{T}, {(\frac{b^3}{b-1} - \eta, 0)}^{T}\right\rbrace$
\end{itemize}
The data point ${\left( 0, 0\right)}^{T}$ will be the root node and each lower layer of the tree will hold the next two outer data points on the line.
$i_{max}$ must be large enough so that the root is able to cover even the most distant data points, i.e. the optimal solution.
\begin{align}
\frac{b^{i_{max} + 1}}{b-1} &> \frac{b^3}{b-1} - \eta \\
\frac{b^{i_{max} + 1}}{b-1} &\geq \frac{b^3}{b-1} \\
i_{max} \geq 2
\end{align}
For simplicity, we assume $i_{max} = 2$ (any layer above $i=2$ will hold only the root).
The layer with $i_{min} = i_{max} - J_{max} - 1$ will hold the whole data set $V$.
Because the distance between any data point in $V$ and ${(0, b + \mu)}^{T}$ is larger than $b^1$, there will be a node for ${(0, b + \mu)}^{T}$ in layer $i=1$.
We will proof the cover tree properties to show that this is a valid cover tree.
(1) Nesting: 
Obviously holds, as $C_{i+1} \subseteq C_i$. 
(2) Covering: 
Holds for $i = 1$, because the maximal distance of any point in $C_1$ to the root ${\left( 0, 0\right)}^{T}$ is $b^2 = b^{i_{max}}$.
Let $1 > i > i_{min}$ be an arbitrary level. 
The nodes ${\left(-\sum_{j=1}^{i_{max} - i-1} b^2 \cdot {\left(\frac{1}{b}\right)}^{j-1},0\right)}^T$ and ${\left(\sum_{j=1}^{i_{max} - i-1} b^2 \cdot {\left(\frac{1}{b}\right)}^{j-1},0\right)}^T$ are the parents of the nodes ${\left(-\sum_{j=1}^{i_{max} - i} b^2 \cdot {\left(\frac{1}{b}\right)}^{j-1},0\right)}^T$ and ${\left(\sum_{j=1}^{i_{max} - i} b^2 \cdot {\left(\frac{1}{b}\right)}^{j-1},0\right)}^T$.
Their distance can be computed as
\begin{align}\label{eq:darbitraryi}
&d\left( {\left(\sum_{j=1}^{i_{max} - i - 1} b^2 \cdot {\left(\frac{1}{b}\right)}^{j-1},0\right)}^T, {\left(\sum_{j=1}^{i_{max} - i} b^2 \cdot {\left(\frac{1}{b}\right)}^{j-1},0\right)}^T \right) \\
&= \sqrt{{\left(  \left( \sum_{j=1}^{i_{max} - i - 1} b^2 \cdot {\left(\frac{1}{b}\right)}^{j-1} \right)- \left( \sum_{j=1}^{i_{max} - i} b^2 \cdot {\left(\frac{1}{b}\right)}^{j-1} \right)  \right)}^2 + 0^2}\\
&= \sqrt{{\left(  \left( \sum_{j=1}^{i_{max} - i - 1} b^2 \cdot {\left(\frac{1}{b}\right)}^{j-1} \right)- \left( \sum_{j=1}^{i_{max} - i - 1} b^2 \cdot {\left(\frac{1}{b}\right)}^{j-1} + b^2 \cdot {\left(\frac{1}{b}\right)}^{i_{max}-i-1}\right)  \right)}^2}\\
&= \sqrt{{\left( - b^2 \cdot {\left(\frac{1}{b}\right)}^{-(i + 1 - i_{max})} \right)}^2}\\
&= \sqrt{{\left( - b^{3-i_{max} + i}\right)}^2}\\
&=  b^{3-i_{max} + i}\\
&=  b^{i+1}
\end{align}
(analogously for negative data points).
Therefore, the covering property holds.
It also holds for level $i_{min}$, because the nodes ${\left(-\sum_{j=1}^{J_{max}} b^2 \cdot {\left(\frac{1}{b}\right)}^{j-1},0\right)}^T$ and ${\left(\sum_{j=1}^{J_{max}} b^2 \cdot {\left(\frac{1}{b}\right)}^{j-1},0\right)}^T$ are the parents of the nodes ${(-\frac{b^3}{b-1} + \eta, 0)}^{T}$ and ${(\frac{b^3}{b-1} - \eta, 0)}^{T}$. Their distance can be computed as
\begin{align} \label{eq:dimin}
&d\left( {\left(\sum_{j=1}^{J_{max}} b^2 \cdot {\left(\frac{1}{b}\right)}^{j-1},0\right)}^T, {\left(\frac{b^3}{b-1} - \eta, 0\right)}^{T} \right) \\
&= \sqrt{{\left(\left( \frac{b^3}{b-1} - \frac{b^{3-J_{max}}}{b-1}\right) - \left( \frac{b^3}{b-1} - \eta \right) \right)}^2 + 0^2}\\
&= \sqrt{{\left( \eta - \frac{b^{3-J_{max}}}{b-1}\right)}^2}\\
&\underset{\leq}{\eqref{eq:eta}} \sqrt{{\left( \frac{b^{3-J_{max}}}{b-1} - b^{i_{min}+1} - \frac{b^{3-J_{max}}}{b-1}\right)}^2}\\
&= b^{i_{min}+1}
\end{align} 
(analogously for the negative data points).
Thus, the covering property holds for every layer of the cover tree.

(3) Separation: 
For every $i_{max} > i \geq i_{min}$, the closest pairwise distance we observe in layer $C_i$ for any node is the distance to the (self-child) of its parent. 
For level $i=1$ the closest distance of nodes in $C_i$ is $b + \mu > b^1$, therefore, the separation property holds. 
Let $1 > i \geq i_{min}$ be an arbitrary level. 
This distance to a parent was computed in Eq. \eqref{eq:darbitraryi} (or Eq. \eqref{eq:dimin}) to be $b^{i + 1}$, which is larger than $b^i$.
Thus, the separation property holds for every layer of the tree.

We have seen, that the above defined cover tree is valid and fullfills the cover tree properties.
The termination layer for $k = 2$ is $C_1$.
When applying \CTRandom, we might get
\begin{equation}
S^{\mathrm{\CTRandom}} = \left\lbrace  {\left( 0, 0 \right)}^T, {\left( 0, b + \mu\right)}^T \right\rbrace
\end{equation}
as a diverse subset, with
\begin{equation}
div( S^{\mathrm{\CTRandom}} ) = b + \mu \approx b.
\end{equation}
Comparing the approximated diverse subset with the optimal solution, we get
\begin{equation}
\underset{\mu, \eta \rightarrow 0}{\lim} \; \; \frac{d^*}{div(S^{\mathrm{\CTRandom}})} = \frac{2b^2}{b-1},
\end{equation}
i.e. the worst possible approximation factor.
Thus, we have shown, that no tighter approximation factor for \CTRandom{} exists.
The following part gives an example with $b$ set to 2.

\subsubsection{Example for $b = 2$}
\begin{figure}
  \centering
  \includegraphics[width=0.9\textwidth]{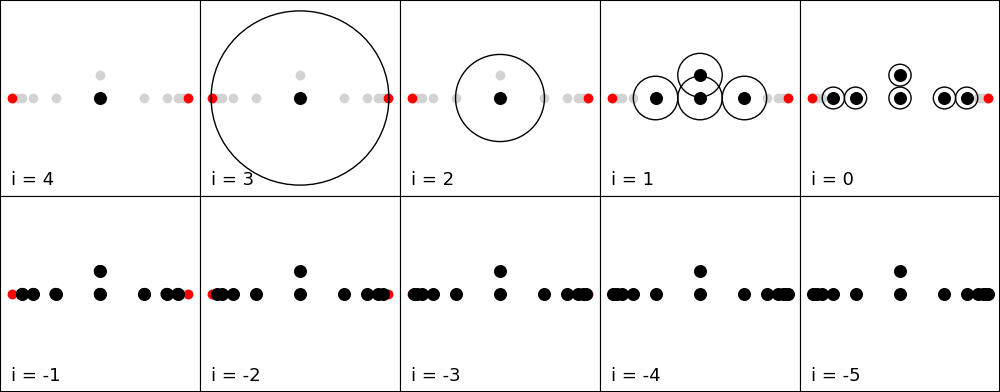}
  \caption{Example for which \CTRandom{} might give a relative diversity of almost 8. Each part of the figure corresponds to one layer of the cover tree. The data set $V$ (with $\mu = \eta = 0.1$) is plotted as grey dots. Nodes are plotted as black dots, encircled with a ball defined by the corresponding covering radius. The optimal solution for $k=2$ with $d^* \approx 16$ is plotted as red dots. \CTRandom{} randomly selects two nodes from the first layer with at least $k$ nodes (here: fourth layer, $i = 1$).}
\label{fig:wcrandom}
\end{figure}

Let the data set be defined as
\begin{equation}
V = \left\lbrace  \left(\begin{array}{c} v\\ 0\\ \end{array} \right), \left(\begin{array}{c} 0\\ 2 + \mu\\ \end{array} \right) \vert v \in \pm \left\lbrace  0, 4, 6, 7, 7.5, 7.75, 7.875, \ldots 8-\eta \right\rbrace \right\rbrace  
\end{equation}
where $\mu, \eta > 0$ small.
Because $\eta \neq 0$, $V$ is finite.
We use Euclidean distance as distance metric.
For $k=2$ the optimal solution is given by ${\left(-(8-\eta), 0\right)}^T$ and ${\left(8-\eta, 0\right)}^T$ with $d^* \approx 16$.
The cover tree is built with $b=2$ such that data points are incrementally added, ordered according to their absolute $x$ value.
Figure \ref{fig:wcrandom} shows the first ten layer of the cover tree with $i_{max} = 4$ (note that the layer $i \geq 2$ only hold the root). \CTRandom{} will randomly select two data points from the first layer with at least $k$ nodes. 
This corresponds to level $i=1$ (fourth layer).
\CTRandom{} might randomly select ${\left(0, 0\right)}^T$ and ${\left(0, 2+\mu\right)}^T$ with $div(S^{\mathrm{\CTRandom}}) \approx 2$.
Thus, we get
\begin{equation}
\underset{\mu, \eta \rightarrow 0}{\lim} \; \; \frac{d^*}{div(S^{\mathrm{\CTRandom}})} = 8
\end{equation}
which shows the tightness of the derived approximation factor.

Note, for this example \CTInherit{} would select a subset with higher diversity: either ${\left(0, 0\right)}^T$ and ${\left(4, 0\right)}^T$, or ${\left(0, 0\right)}^T$ and ${\left(-4, 0\right)}^T$ will be the selected nodes. Thus, $div(S^{\mathrm{\CTInherit}}) = 4$.

\subsection{Tightness of $\alpha^{\mathrm{\CTGreedy}}$ and $\alpha^{\mathrm{\CTInherit}}$}\label{sec:thightnessInherit}
Let the data set be defined as
\begin{equation}
V = \left\lbrace  v \vert v \in \left\lbrace 0, \pm \left( 1 - \mu \right), \pm \left( \frac{b}{b-1} + 1 - \eta \right) \right\rbrace \; \; \cup \bigcup_{J \in [1, J_{max}]} \pm \left\lbrace 1 + \sum_{j=1}^{J} {\frac{1}{b}}^{j-1} \right\rbrace \right\rbrace,  
\end{equation}
where $\mu, \eta > 0$ small.
Note that, the value of $\mu$ must be chosen such that $i = 0$ is the largest possible integer with $b^i < 1 + \mu$.
Furthermore, $J_{max}$ is the smallest integer such that
\begin{equation}\label{eq:eta_greedy}
{\frac{1}{b}}^{J_{max}} \leq \eta.
\end{equation}
All of the data points in $V$ lie on the line between $-\left( \frac{b}{b-1} + 1 - \mu \right)$ and $\left( \frac{b}{b-1} + 1 - \mu \right)$, where the distance between neighboring data points decreases as we move to the ends of the line.
We choose the Euclidean distance as distance metric.
The optimal solution $S^* \subseteq V$ for $k=2$ is given by 
\begin{equation}
S^* = \left\lbrace -\left( \frac{b}{b-1} + 1 - \eta \right) , \left( \frac{b}{b-1} + 1 - \eta \right) \right\rbrace
\end{equation}
with a diversity of
\begin{equation}
d^* = 2 \left( \frac{b}{b-1} + 1 \right) - 2 \eta \;\; \approx \;\; 2 \left( \frac{b}{b-1} + 1\right).
\end{equation}
The cover tree for $V$ is built with base $b$ such that data points are incrementally added, ordered according to their absolute $x$ value.
As a result, we get the following cover tree:
\begin{itemize}
\item $i > i_{ter}$: $C_{i} = \left\lbrace 0 \right\rbrace$
\item $i_{ter} = 0$: $C_{i_{ter}} = C_{i_{ter}+1} \cup \left\lbrace - \left( 1 + \mu \right) , 1 + \mu \right\rbrace$
\item $i_{ter}>i>i_{min}$: $C_i = C_{i+1} \cup \left\lbrace - \left( \sum_{j=1}^{-i} {\frac{1}{b}}^{j-1} \right) ,  \sum_{j=1}^{-i} {\frac{1}{b}}^{j-1} \right\rbrace$
\item $i_{min}$: $C_{i_{min}} = C_{i_{min} + 1} \cup \left\lbrace -\left( \frac{b}{b-1} + 1 - \eta \right) ,  \frac{b}{b-1} + 1 - \eta \right\rbrace$.
\end{itemize}
The data point $0$ will be the root node and each lower layer of the tree will hold the next two outer data points on the line.
Any layer before level $i_{ter}$ (the termination layer) will hold only the root.
The layer with $i_{min} = - \left( J_{max} + 1 \right)$ will hold the whole data set $V$.
We will proof the cover tree properties to show that this is a valid cover tree.
(1) Nesting: 
Obviously holds, as $C_{i+1} \subseteq C_i$. 
(2) Covering: 
Obviously holds for $i > i_{ter}$, because those layers only hold the root.
For $i = i_{ter} = 0$, the maximal distance of any point in $C_0$ to the root $0$ is $d(0, 1 + \mu) = 1 + \mu$.
Since $i_{ter}$ is the largest possible integer $i$, such that $b^i < 1 + \mu$ (see above), it follows $b^{i_{ter} + 1} \geq 1 + \mu$.
Let $i_{ter} > i > i_{min}$ be an arbitrary level. 
The nodes $ - \left( 1 + \sum_{j=1}^{-(i+1)} {\frac{1}{b}}^{j-1} \right)$ and $1 + \sum_{j=1}^{-(i+1)} {\frac{1}{b}}^{j-1}$ are the parents of the nodes $1 + \sum_{j=1}^{-i} {\frac{1}{b}}^{j-1}$ and $1 + \sum_{j=1}^{-i} {\frac{1}{b}}^{j-1}$.
Their distance can be computed as
\begin{align}\label{eq:darbitraryi_greedy}
&d\left( 1 + \sum_{j=1}^{-(i+1)} {\frac{1}{b}}^{j-1} , 1 + \sum_{j=1}^{-i} {\frac{1}{b}}^{j-1} \right) \\
&= \sqrt{{\left(\left( 1 + \sum_{j=1}^{-(i+1)} {\frac{1}{b}}^{j-1}\right) - \left(1 + \sum_{j=1}^{-i} {\frac{1}{b}}^{j-1}\right)\right)}^2}\\
&= \sqrt{{\left(\left( 1 + \sum_{j=1}^{-(i+1)} {\frac{1}{b}}^{j-1} \right) - \left(1 + \sum_{j=1}^{-(i+1)} {\frac{1}{b}}^{j-1} + {\frac{1}{b}}^{-(i+1)}\right)\right)}^2}\\
&= {\frac{1}{b}}^{-(i+1)}\\
&= {b}^{i+1}
\end{align}
(analogously for negative data points).
Therefore, the covering property holds.
It also holds for level $i_{min}$, because the nodes $-\left(1 + \sum_{j=1}^{J_{max}}{\frac{1}{b}}^{j-1}\right)$ and $1 + \sum_{j=1}^{J_{max}}{\frac{1}{b}}^{j-1}$ are the parents of the nodes $- \left( \frac{b}{b-1} + 1 - \eta\right)$ and $ \frac{b}{b-1} + 1 - \eta$. Their distance can be computed as
\begin{align} \label{eq:dimin_greedy}
&d\left( 1 + \sum_{j=1}^{J_{max}}{\frac{1}{b}}^{j-1}, \frac{b}{b-1} + 1 - \eta \right) \\
&= \sqrt{{\left(\left( \frac{b}{b-1} + 1 - \frac{{\frac{1}{b}}^{J_{max} - 1}}{b-1} \right) - \left( \frac{b}{b-1} + 1 - \eta \right) \right)}^2}\\
&= \sqrt{{\left( \eta - \frac{{\frac{1}{b}}^{J_{max} - 1}}{b-1} \right)}^2}\\
&\underset{\leq}{\eqref{eq:eta_greedy}} \sqrt{{\left( {\frac{1}{b}}^{J_{max}} - \frac{{\frac{1}{b}}^{J_{max} - 1}}{b-1} \right)}^2}\\
&= \sqrt{{\left( {\frac{1}{b}}^{J_{max}} \left( 1 - b\right) \left( \frac{1}{b-1}\right)\right)}^2}\\
&= \sqrt{{\left( -{\frac{1}{b}}^{J_{max}} \right)}^2}\\
&= {\frac{1}{b}}^{J_{max}}\\
&= {\frac{1}{b}}^{-(i_{min} + 1)}\\
&= b^{i_{min} + 1}\\
\end{align} 
(analogously for the negative data points).
Thus, the covering property holds for every layer of the cover tree.

(3) Separation: 
For every $i_{max} > i \geq i_{min}$, the closest pairwise distance we observe in layer $C_i$ for any node is the distance to the (self-child) of its parent. 
For level $i=i_{ter}=0$ the closest distance of nodes in $C_i$ is $1 + \mu > b^0$, therefore, the separation property holds. 
Let $0 > i \geq i_{min}$ be an arbitrary level. 
This distance to a parent was computed in Eq. \eqref{eq:darbitraryi_greedy} (or Eq. \eqref{eq:dimin_greedy}) to be $b^{i + 1}$, which is larger than $b^i$.
Thus, the separation property holds for every layer of the tree.

We have seen, that the above defined cover tree is valid and fullfills the cover tree properties.
The termination layer for $k = 2$ is $C_0$.
When applying \CTInherit, we will get
$S^{\mathrm{\CTInherit}} = \left\lbrace  0, 1 + \mu \right\rbrace$ or $S^{\mathrm{\CTInherit}} = \left\lbrace  0, - \left(1 + \mu\right) \right\rbrace$
as a diverse subset, with
\begin{equation}
div( S^{\mathrm{\CTInherit}} ) = 1 + \mu \approx 1.
\end{equation}
Comparing the approximated diverse subset with the optimal solution, we get
\begin{equation}
\underset{\mu, \eta \rightarrow 0}{\lim} \; \; \frac{d^*}{div(S^{\mathrm{\CTInherit}})} = \frac{2b}{b-1} + 2,
\end{equation}
i.e. the worst possible approximation factor.
\CTGreedy{} can give the same solution when the root node is randomly chosen in the first iteration.
Thus, we also get
\begin{equation}
\underset{\mu, \eta \rightarrow 0}{\lim} \; \; \frac{d^*}{div(S^{\mathrm{\CTGreedy}})} = \frac{2b}{b-1} + 2.
\end{equation}
Thus, we have shown, that no tighter approximation factor for \CTGreedy{} and \CTInherit{} exists.
The following part gives an example with $b$ set to 2.

\subsubsection{Example for $b=2$}
\begin{figure}
  \centering
  \includegraphics[width=0.9\textwidth]{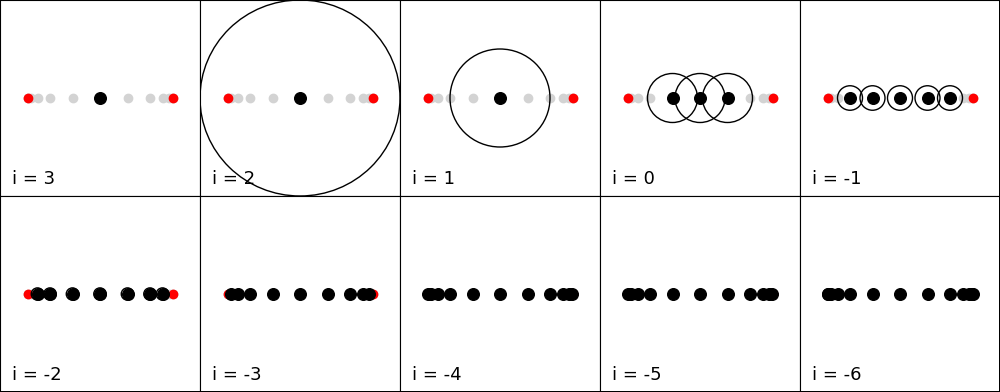}
  \caption{Example for which \CTGreedy{} and \CTInherit{} might give a relative diversity of almost 6. Each part of the figure corresponds to one layer of the cover tree. The data set $V$ (with $\mu = \eta = 0.1$) is plotted as gray dots. Nodes are plotted as black dots, encircled with a ball defined by the corresponding covering radius. The optimal solution for $k=2$ with $d^* \approx 6$ is plotted as red dots. \CTInherit{} selects the root node and one of the two nodes from the first layer with at least $k$ nodes (here: fourth layer, $i = 0$). \CTGreedy{} can return the same solution as \CTInherit, when the root is selected randomly as first data point of the solution.}
\label{fig:wcgreedy}
\end{figure}

Let the data set be defined as
\begin{equation}
V = \left\lbrace  v \vert v \in \pm \left\lbrace  0, 1+\mu, 2, 2.5, 2.75, 2.875, \ldots 3-\eta \right\rbrace \right\rbrace  
\end{equation}
where $\mu, \eta > 0$ small.
Because $\eta \neq 0$, $V$ is finite.
We use Euclidean distance as distance metric.
For $k=2$ the optimal solution is given by $-(3-\eta)$ and $3-\eta$ with $d^* \approx 6$.
The cover tree is built with $b=2$ such that data points are incrementally added, ordered according to their absolute value.
Figure \ref{fig:wcgreedy} shows the first ten layer of the cover tree with $i_{max} = 3$. The first layer with at least $k$ nodes is level $i=0$ (fourth layer).
\CTInherit{} is initialized with the root and selects one of the remaining nodes of level $0$.
Thus, $S^{\mathrm{\CTInherit}} = \lbrace 0, -(1+\mu)\rbrace$ or $S^{\mathrm{\CTInherit}} = \lbrace 0, 1+\mu \rbrace$ with $div(S^{\mathrm{\CTInherit}}) \approx 1$.
Thus, we get
\begin{equation}
\underset{\mu, \eta \rightarrow 0}{\lim} \; \; \frac{d^*}{div(S^{\mathrm{\CTInherit}})} = 6
\end{equation}
which shows the tightness of the derived approximation factor.

When the root is selected randomly as first data point of the solution, \CTGreedy{} can return the same solution as \CTInherit.
Thus, we also get
\begin{equation}
\underset{\mu, \eta \rightarrow 0}{\lim} \; \; \frac{d^*}{div(S^{\mathrm{\CTGreedy}})} = 6.
\end{equation}

\section{Additional Figures and Tables}\label{sec:additionalfigures}

\begin{figure}[t]
  \centering
  \includegraphics[width=0.95\textwidth]{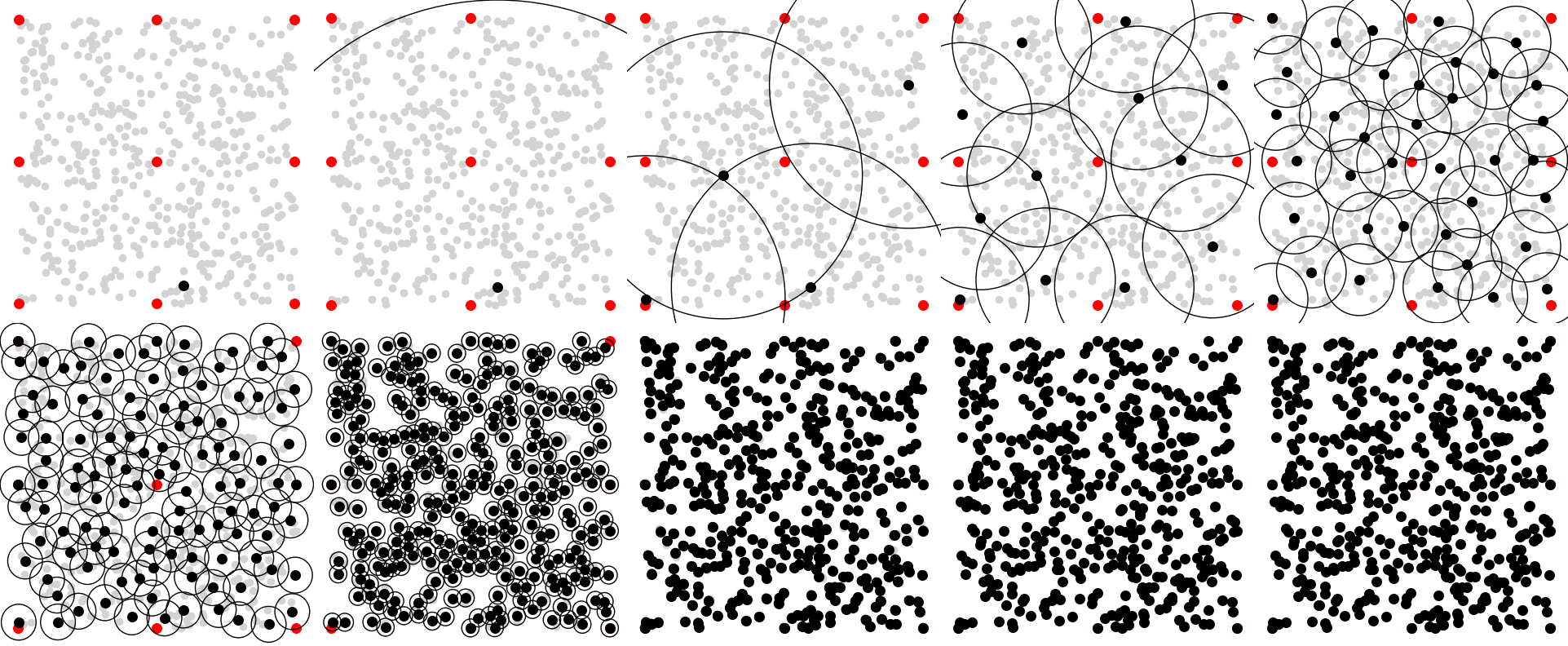}
  \caption{Example of a cover tree with $b=2$ for a two dimensional data set (grey, red). Each plot shows one level $i$ of the cover tree with all nodes (black) encircled with a ball of radius $2^i$. The red dots correspond to the optimal diverse solution with $k = 9$.}
\label{fig:covertree}
\end{figure}

Figure \ref{fig:covertree} shows each layer of a cover tree built with $b=2$ on a 2D grid data example with $k=9$.

 %We conducted various experiments on artificial as well as on real data sets. 
 Table \ref{table:experiments} summarizes information for the artificial and real data in our experiments. 
 For the cover tree we used the implementation provided in \cite{drosou2013poikilo}.

\begin{table}[t]
\centering
\caption{Summary of conducted experiments. }
\label{table:experiments}
\begin{tabular}{llllll}
\toprule
& Experiment & Dimension & Sample Size     & Diverse Set Size k & Distance Metric \\ 
\midrule
\multirow{2}{*}{Artificial} 
& Grid 2D    & $2$         & $\lbrace500, 1000, 5000\rbrace$ & $\lbrace4,9,25\rbrace$             & Euclidean       \\
& Grid 5D    & $5$         & $5000$            & $\lbrace32,243,1024\rbrace$        & Euclidean       \\ 
\midrule
\multirow{3}{*}{Real}
& Cities     & $2$         & $10975$           & $[2, 100]$       & Euclidean       \\
& Faces      & $624$       & $3840$            & $[2, 100]$        & Cosine          \\
& MNIST      & $784$       & $70000$           & $[2, 100]$        & Cosine          \\ 
\bottomrule
\end{tabular}
\end{table}

\begin{figure}[t]
  \centering
  \includegraphics[width=0.95\textwidth]{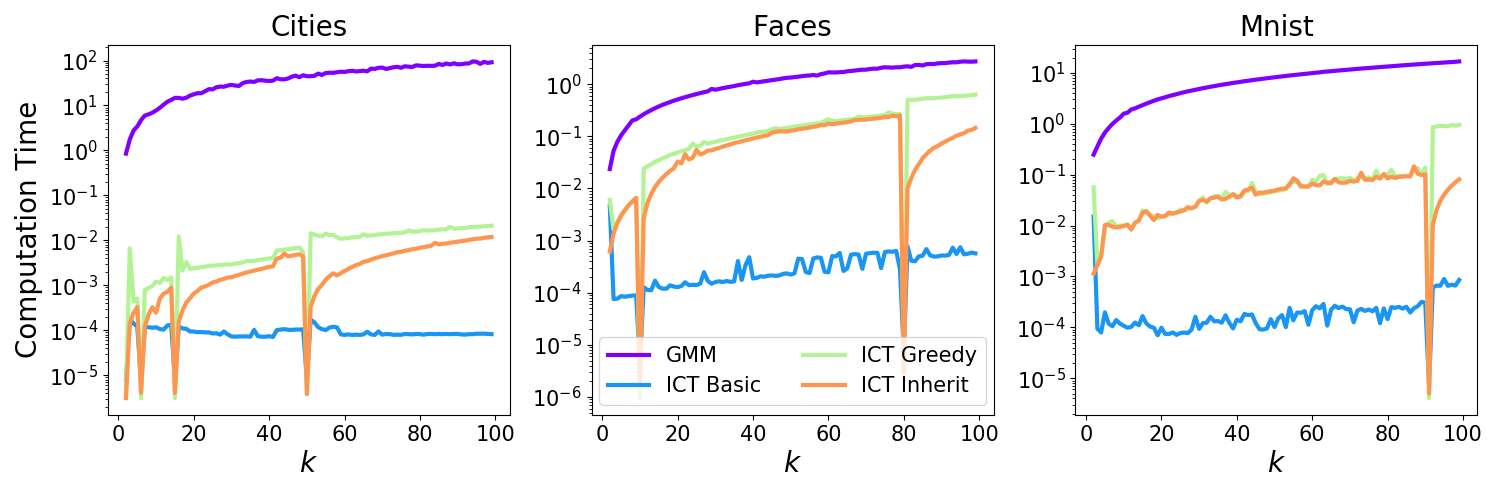}
  \caption{Computation time in ms on the \cities, \faces{} and \mnist{} data sets.
  All experiments were conducted on a Intel(R) Xeon(R) CPU E5-2640 v4 with 2.40GHz.}
\label{fig:Af_Ct}
\end{figure}

Figure \ref{fig:Af_Ct} shows the computation time of selecting the diverse subsets for the different approaches.
One can clearly see the efficiency of the cover-tree approaches.
Compared to GMM, \CTRandom, \CTGreedy{} and \CTInherit{} have fast computation time even for large $k$.
The computation time of the cover tree approaches shows a step function behavior.
Again, this can be explained by the layer-wise structure of the cover tree. 
When the termination layer holds exactly $k$ nodes, no selection must be made. 
We can also see the difference in the complexity of \CTGreedy{} and \CTInherit. 
When the layer before the termination layer holds almost $k$ nodes, i.e. $\vert C_{i+1} \vert \approx \vert C_i \vert$, \CTInherit{} only has to select few nodes. 

\ifOmitHistogramSmallB
\else
Figure \ref{fig:distAFB12} and \ref{fig:distAFB16} show the approximated distribution of the relative inverse diversity on the 2D grid data for cover trees with different bases ($b = 1.2$ and $b=1.6$).
\begin{figure}
  \centering
  \includegraphics[width=0.7\textwidth]{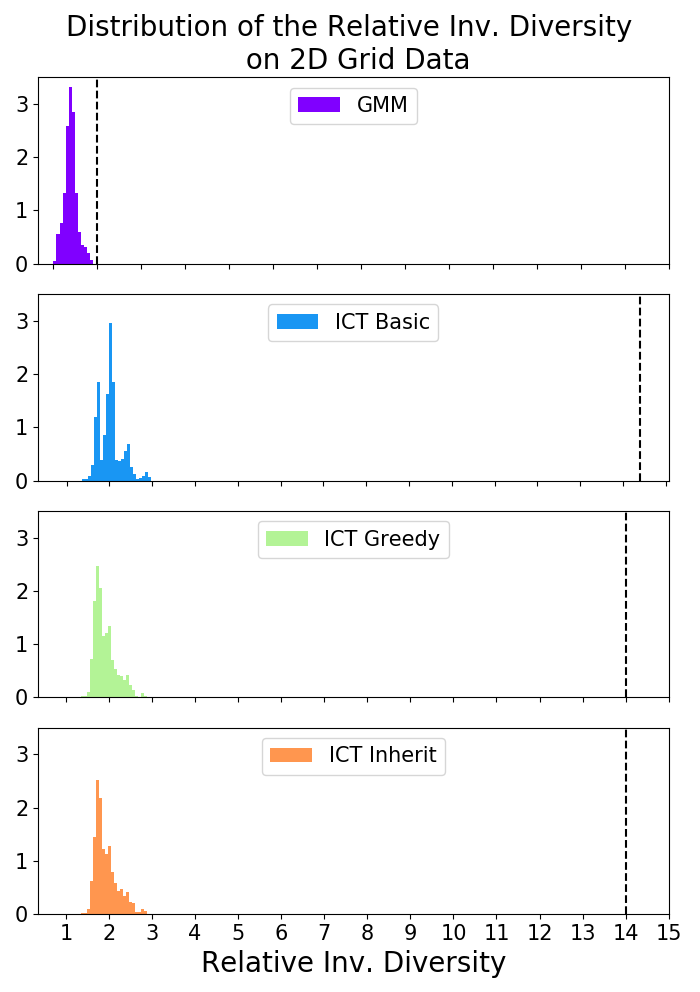}
  \caption{Estimated distribution of the relative inverse diversity on the 2D grid data. The cover tree was built with $b=1.2$. The dashed vertical lines indicate the approximation factor.}
\label{fig:distAFB12}
\end{figure}

% For the sheep set, the MCMC selects three images with very similar background which only differ due to small dispositions of the contained sheep. 
% For the k=5 diverse sheep images of our method, the images with similar background show different sized and positioned sheep.
% For the volcano images, the diverse set contains a set of more diverse colors of the sky in the background as well as more diverse volcanic peaks as compared to the converged MCMC set.
% This trend for selecting diverse set of colors can be also seen in the set of greenhouses selected by MCMC which shows two pairs of similar images. 
% In contrast, our CV set contains five quite different generated greenhouses regarding the colors of the plants and sky. hile 

\begin{figure}
  \centering
  \includegraphics[width=0.7\textwidth]{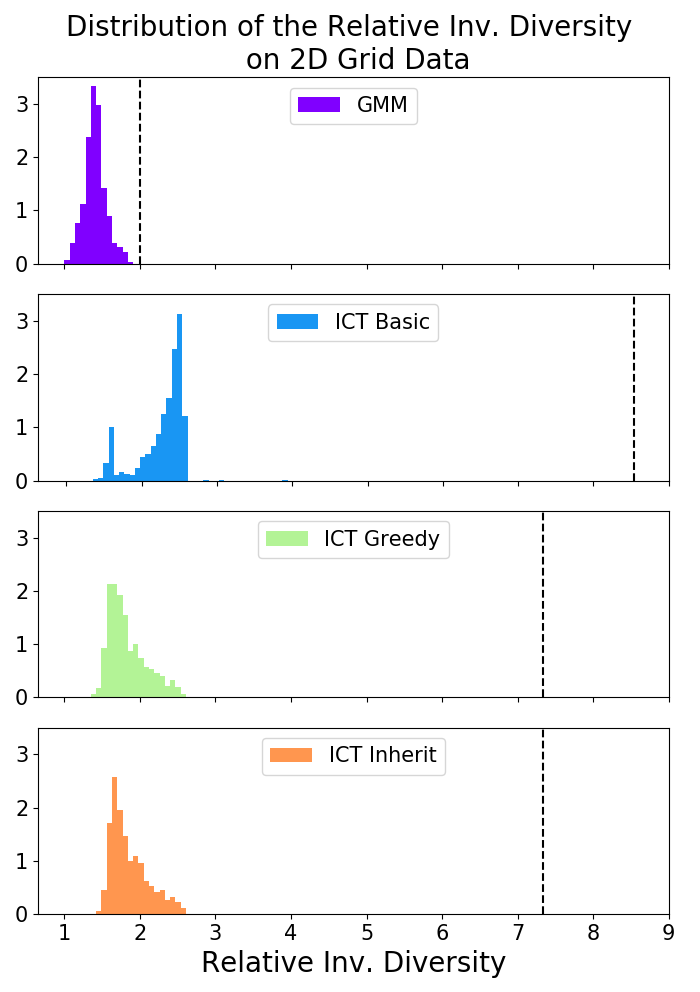}
  \caption{Estimated distribution of the relative inverse diversity on the 2D grid data. The cover tree was built with $b=1.6$. The dashed vertical lines indicate the approximation factor.}
  \label{fig:distAFB16}
\end{figure}

\fi

\section{Examples of diverse images generated by PPGN}
\label{sec:MoreExamplesPPGN}
\begin{figure}[h]
\centering
\begin{subfigure}{0.45\textwidth}
\centering
\includegraphics[width=0.95\textwidth]{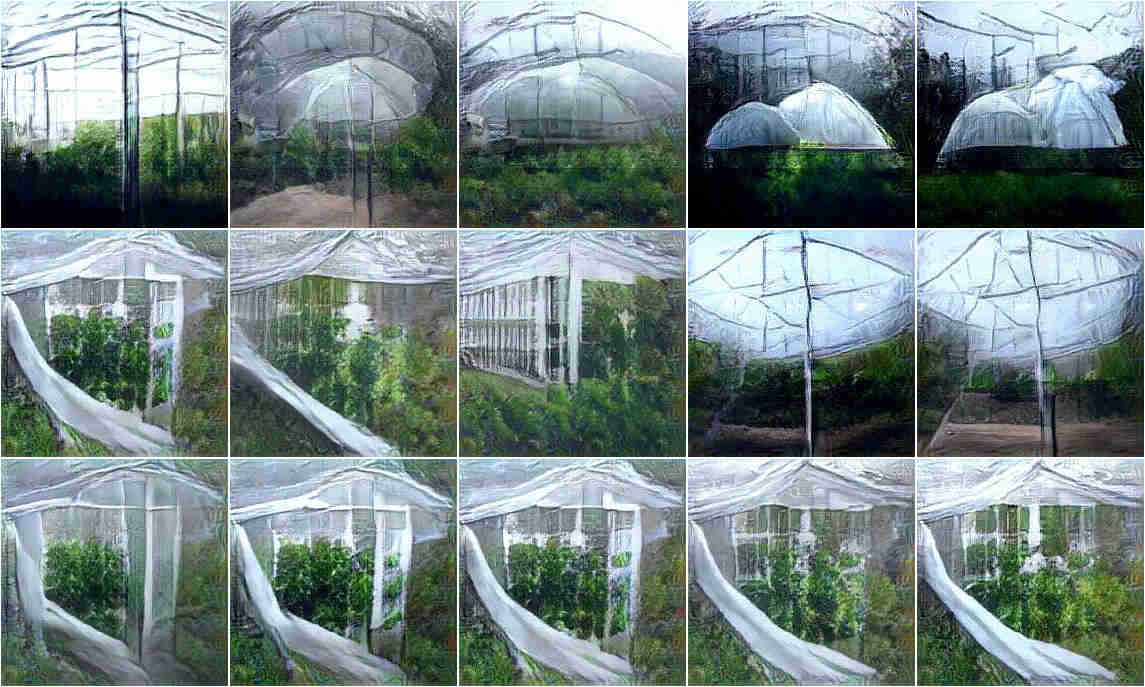}
\end{subfigure}
\centering
\begin{subfigure}{0.45\textwidth}
\includegraphics[width=0.95\textwidth]{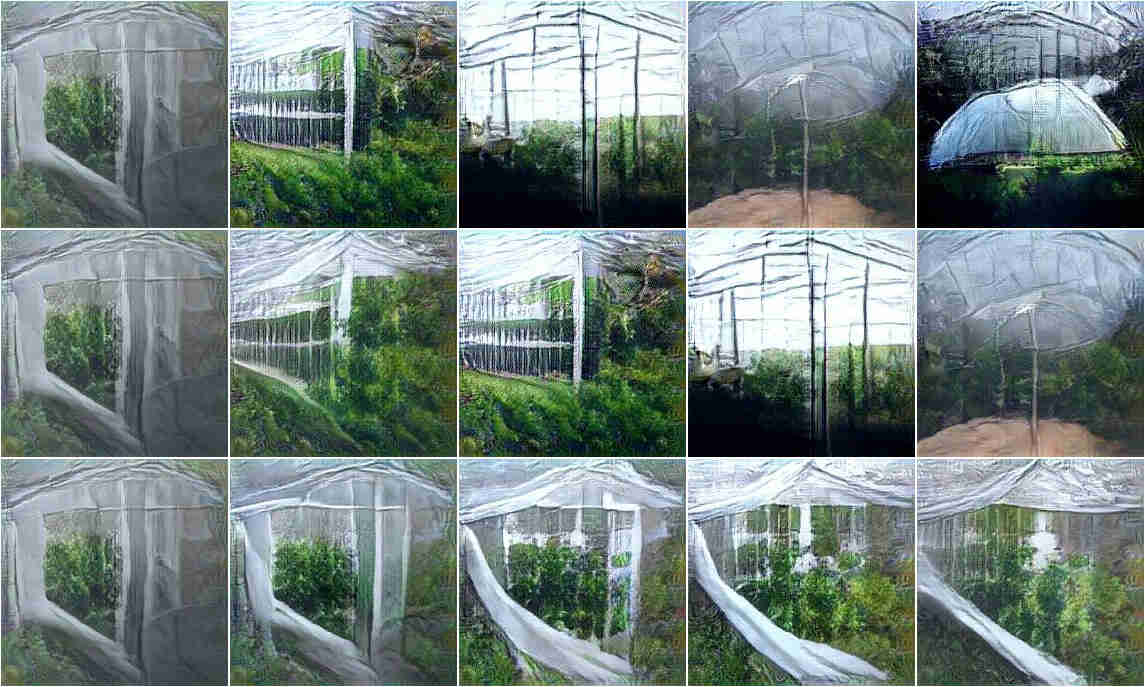}
\end{subfigure}
\caption{$k=5$ randomly chosen images (left) from the MCMC sequence of PPGN
for \emph{Greenhouse} class, and the corresponding  diverse sets (right) chosen by \CTInherit,
after 20 (bottom), 100 (middle), and 200 (top) MCMC steps.}
\label{fig:greenhouse}
\end{figure}

\begin{figure}[h]
\centering
\begin{subfigure}{0.45\textwidth}
\centering
\includegraphics[width=0.95\textwidth]{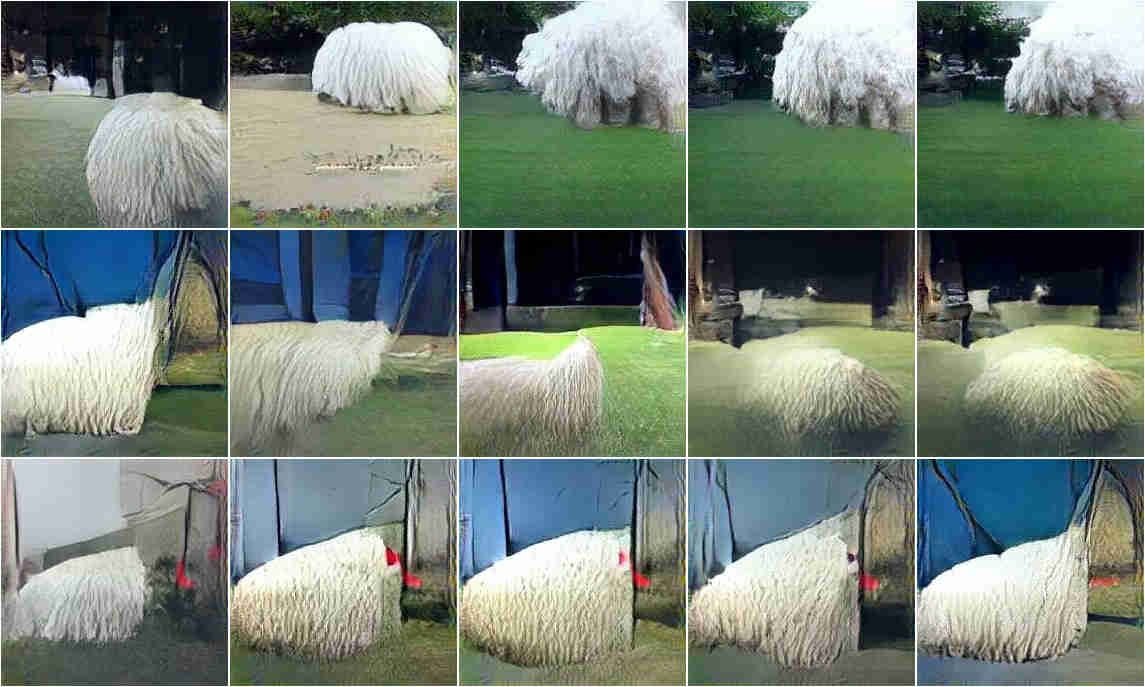}
\end{subfigure}
\centering
\begin{subfigure}{0.45\textwidth}
\includegraphics[width=0.95\textwidth]{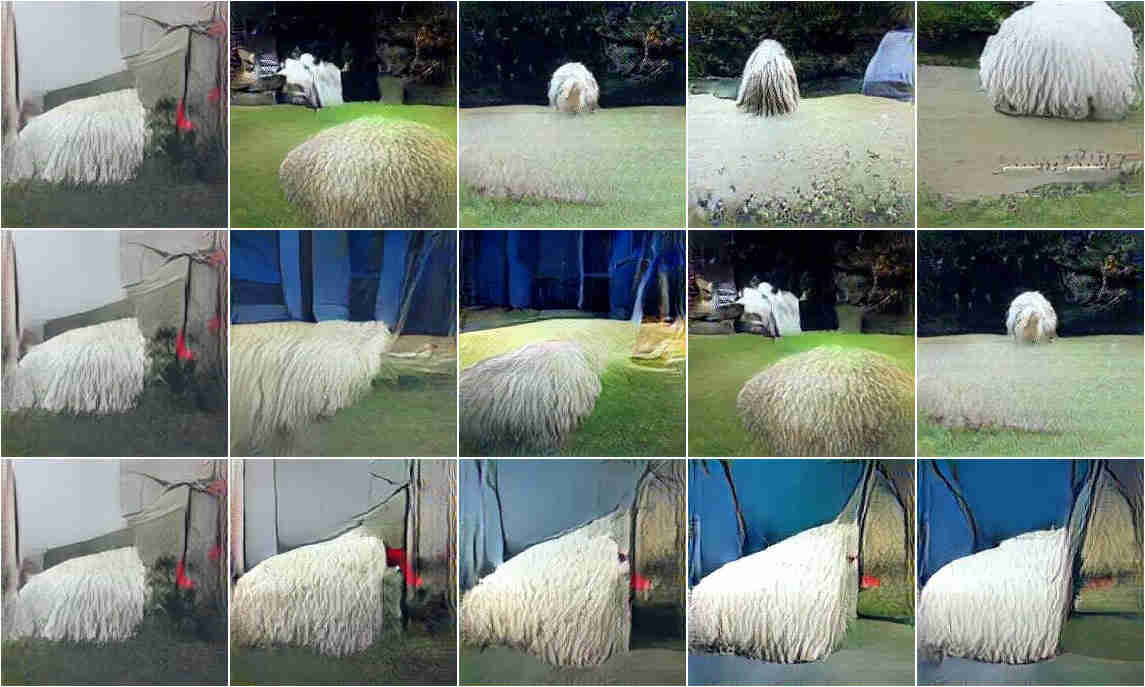}
\end{subfigure}
\caption{$k=5$ randomly chosen images (left) from the MCMC sequence of PPGN
for \emph{Sheep} class, and the corresponding  diverse sets (right) chosen by \CTInherit,
after 20 (bottom), 100 (middle), and 200 (top) MCMC steps.}
\label{fig:sheep}
\end{figure}

\begin{figure}[h]
\centering
\begin{subfigure}{0.45\textwidth}
\centering
\includegraphics[width=0.95\textwidth]{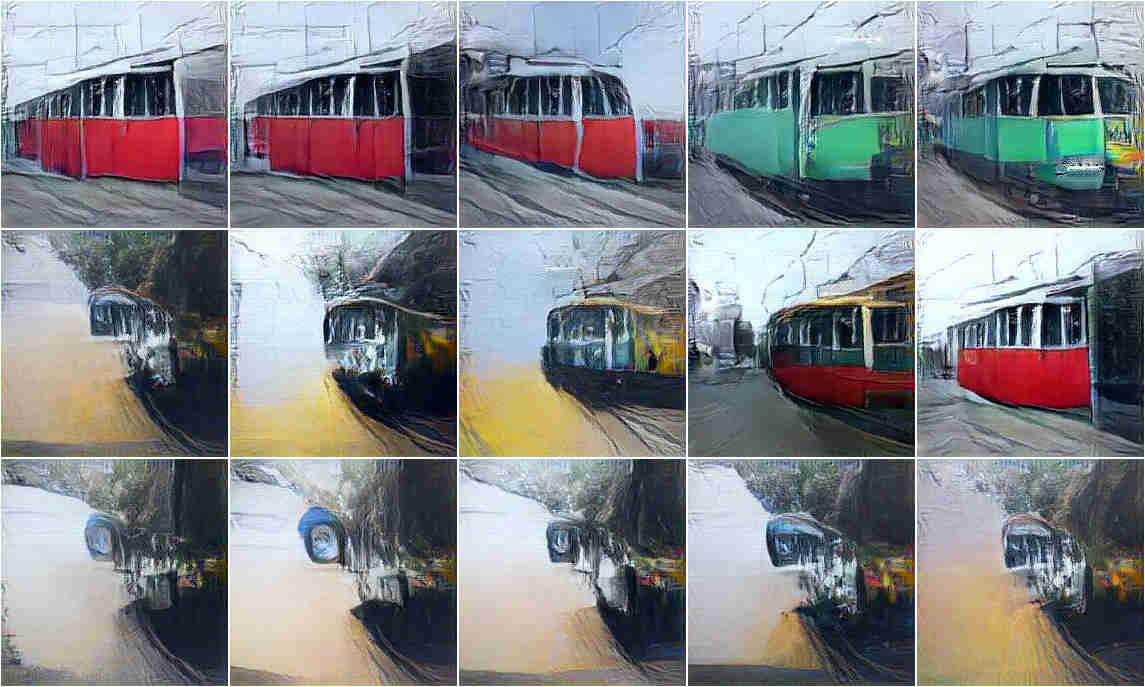}
\end{subfigure}
\centering
\begin{subfigure}{0.45\textwidth}
\includegraphics[width=0.95\textwidth]{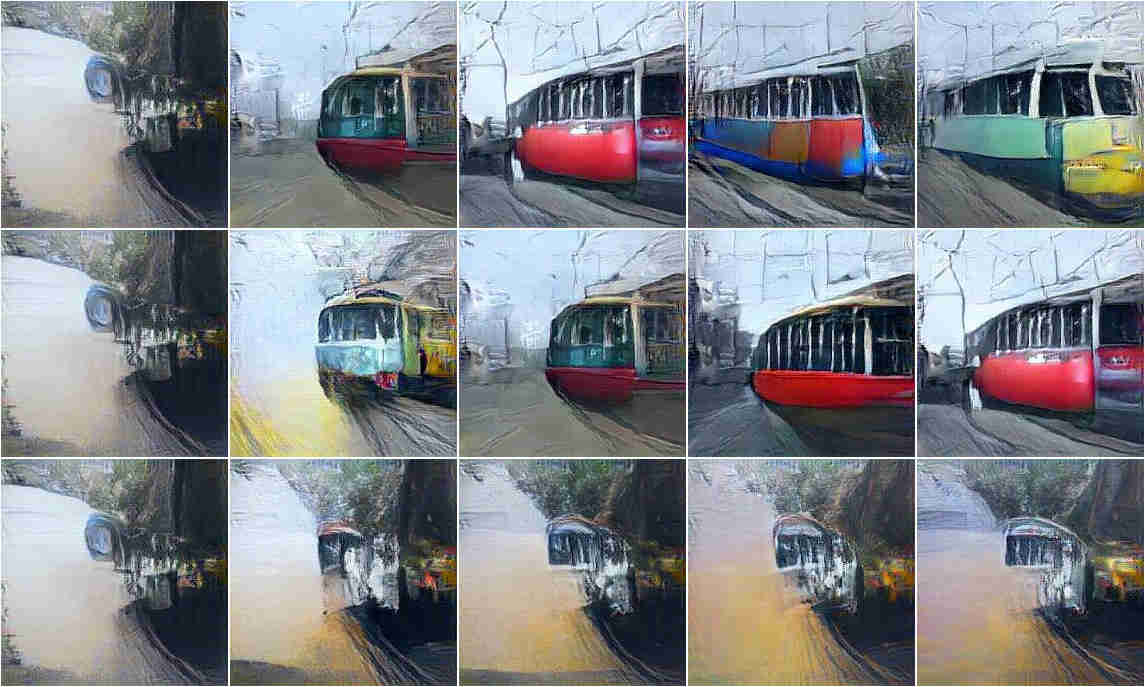}
\end{subfigure}
\caption{$k=5$ randomly chosen images (left) from the MCMC sequence of PPGN
for \emph{Train} class, and the corresponding  diverse sets (right) chosen by \CTInherit,
after 20 (bottom), 100 (middle), and 200 (top) MCMC steps.}
\label{fig:train}
\end{figure}

\begin{figure}[h]
\centering
\begin{subfigure}{0.45\textwidth}
\centering
\includegraphics[width=0.95\textwidth]{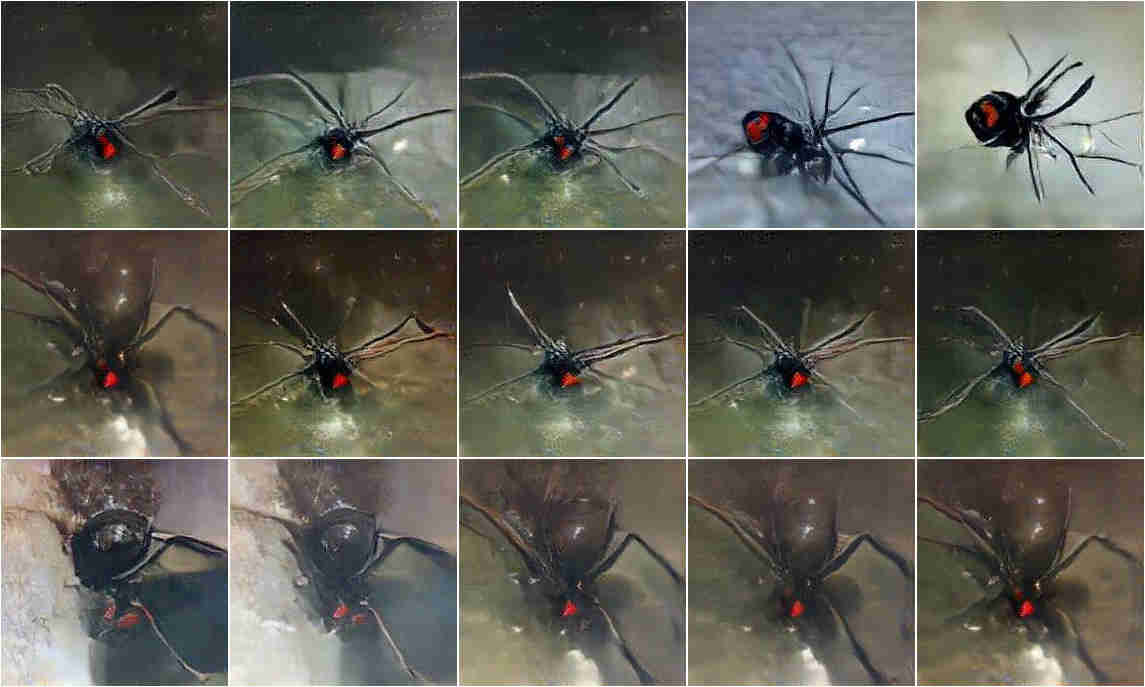}
\end{subfigure}
\centering
\begin{subfigure}{0.45\textwidth}
\includegraphics[width=0.95\textwidth]{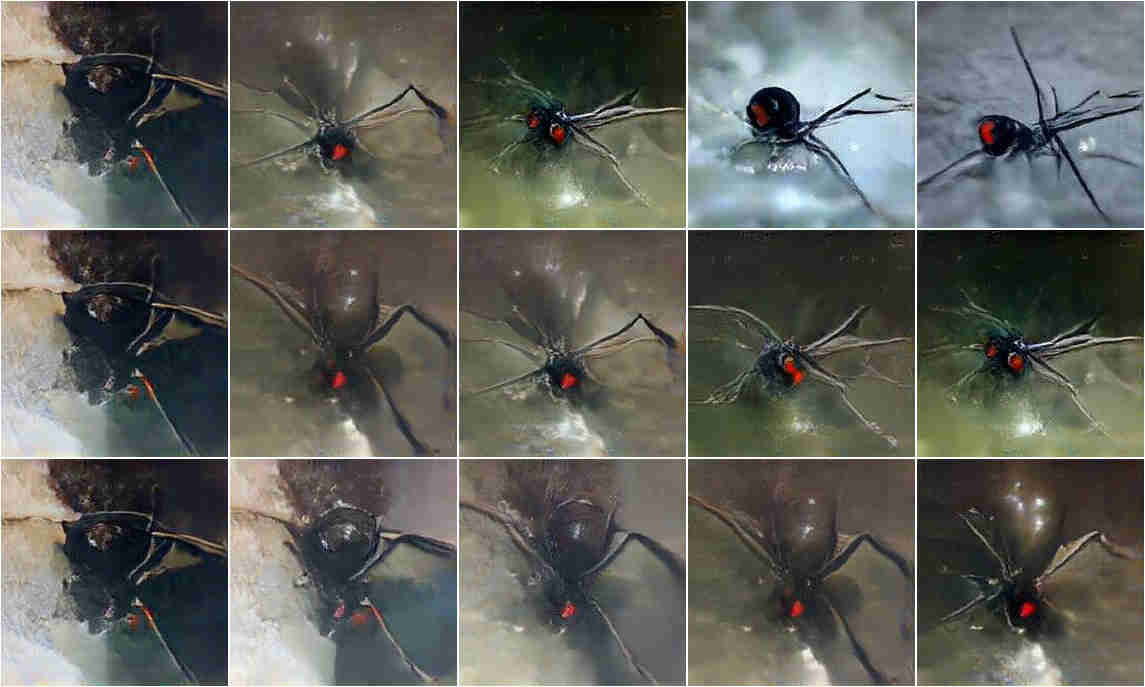}
\end{subfigure}
\caption{$k=5$ randomly chosen images (left) from the MCMC sequence of PPGN
for \emph{Spider} class, and the corresponding  diverse sets (right) chosen by \CTInherit,
after 20 (bottom), 100 (middle), and 200 (top) MCMC steps.}
\label{fig:spider}
\end{figure}

\begin{figure}[h]
\centering
\begin{subfigure}{0.45\textwidth}
\centering
\includegraphics[width=0.95\textwidth]{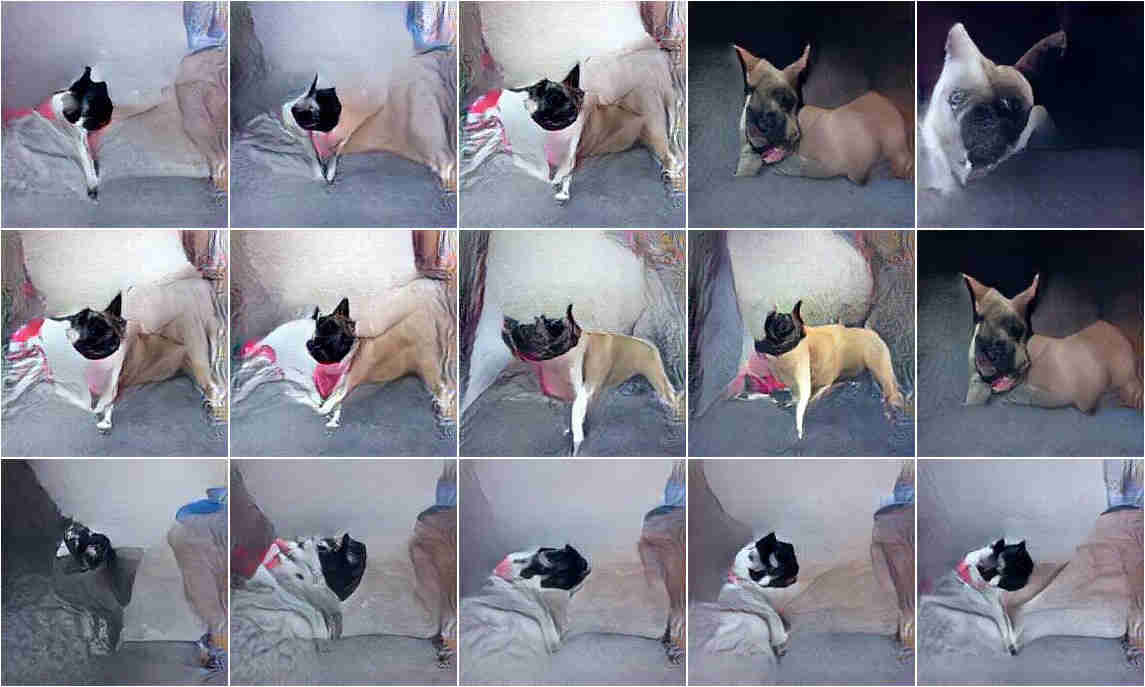}
\end{subfigure}
\centering
\begin{subfigure}{0.45\textwidth}
\includegraphics[width=0.95\textwidth]{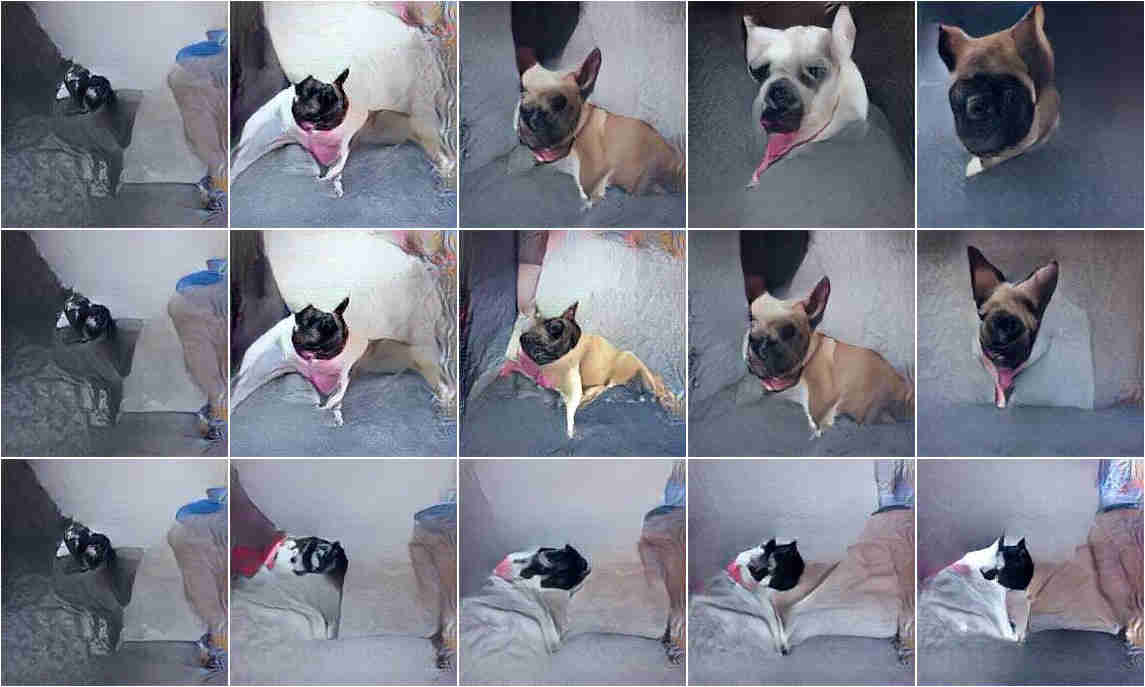}
\end{subfigure}
\caption{$k=5$ randomly chosen images (left) from the MCMC sequence of PPGN
for \emph{Dog} class, and the corresponding  diverse sets (right) chosen by \CTInherit,
after 20 (bottom), 100 (middle), and 200 (top) MCMC steps.}
\label{fig:dog}
\end{figure}

\begin{figure}[h]
\centering
\begin{subfigure}{0.45\textwidth}
\centering
\includegraphics[width=0.95\textwidth]{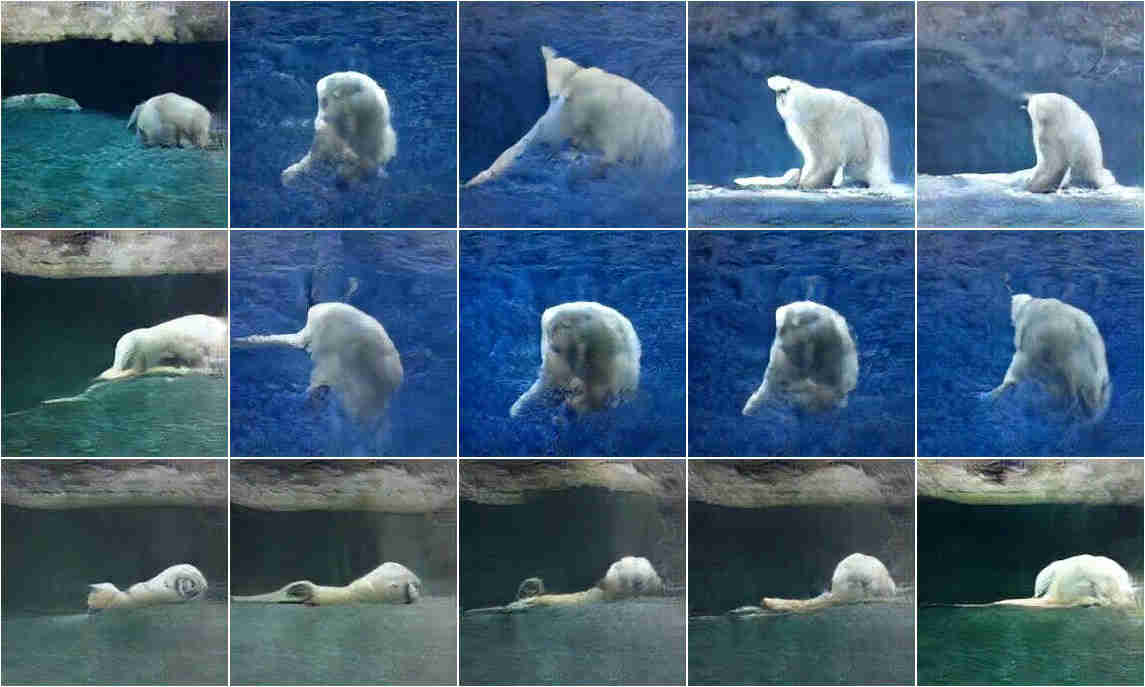}
\end{subfigure}
\centering
\begin{subfigure}{0.45\textwidth}
\includegraphics[width=0.95\textwidth]{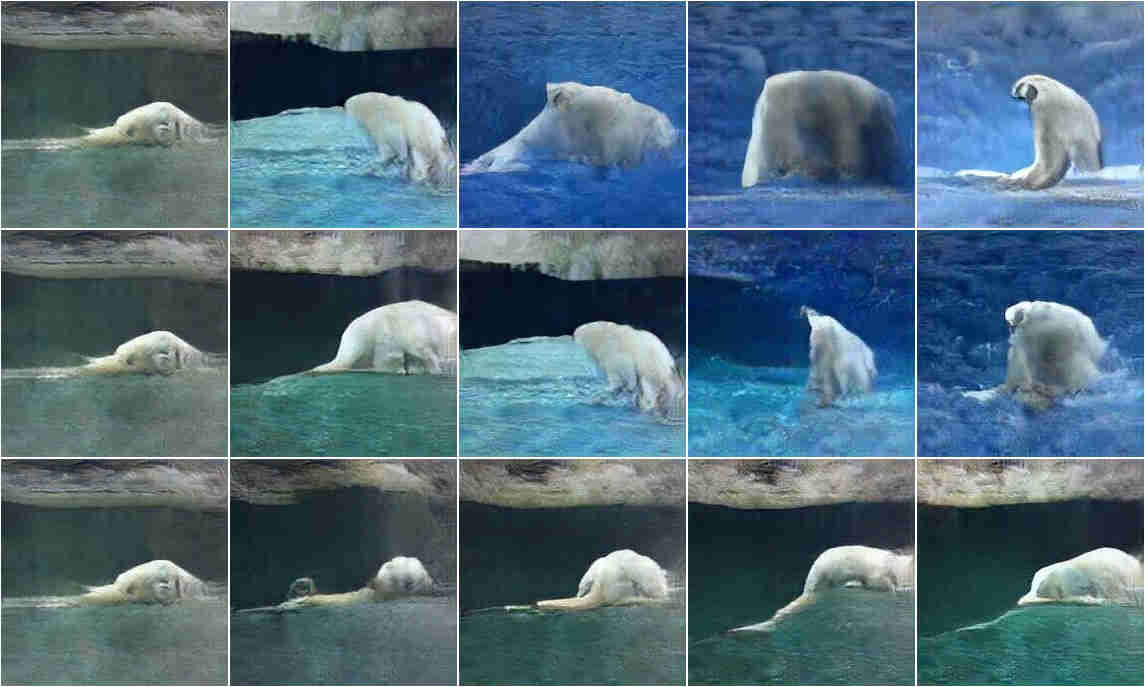}
\end{subfigure}
\caption{$k=5$ randomly chosen images (left) from the MCMC sequence of PPGN
for \emph{Polar Bear} class, and the corresponding  diverse sets (right) chosen by \CTInherit,
after 20 (bottom), 100 (middle), and 200 (top) MCMC steps.}
\label{fig:polarbear}
\end{figure}

\begin{figure}[h]
\centering
\begin{subfigure}{0.45\textwidth}
\centering
\includegraphics[width=0.95\textwidth]{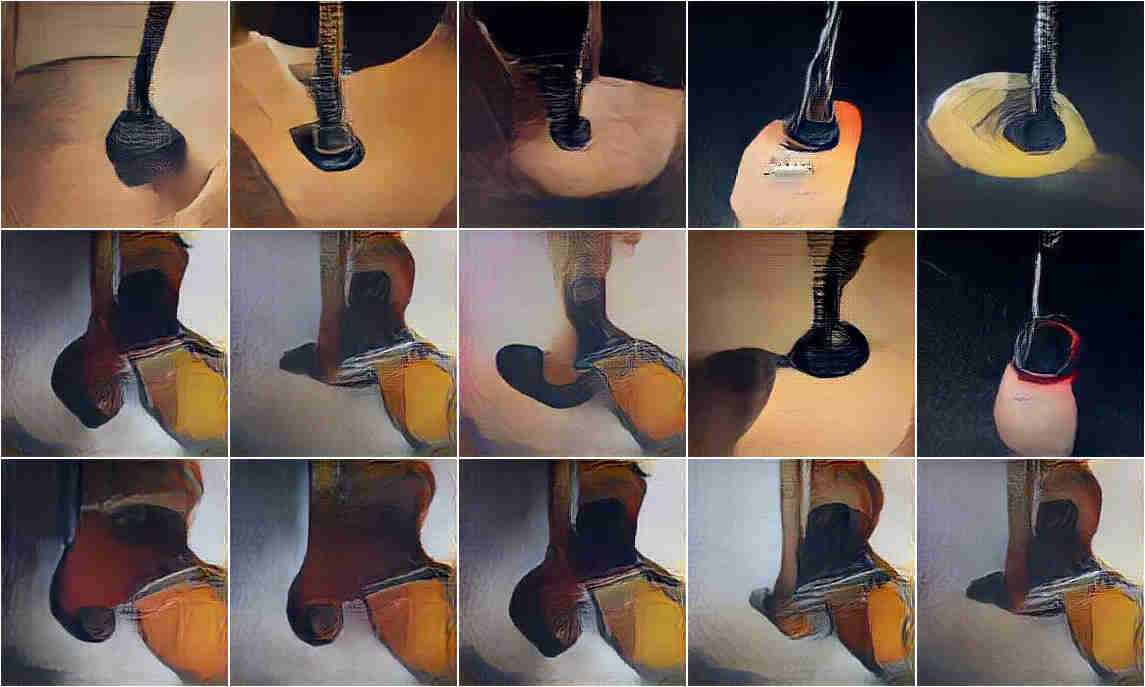}
\end{subfigure}
\centering
\begin{subfigure}{0.45\textwidth}
\includegraphics[width=0.95\textwidth]{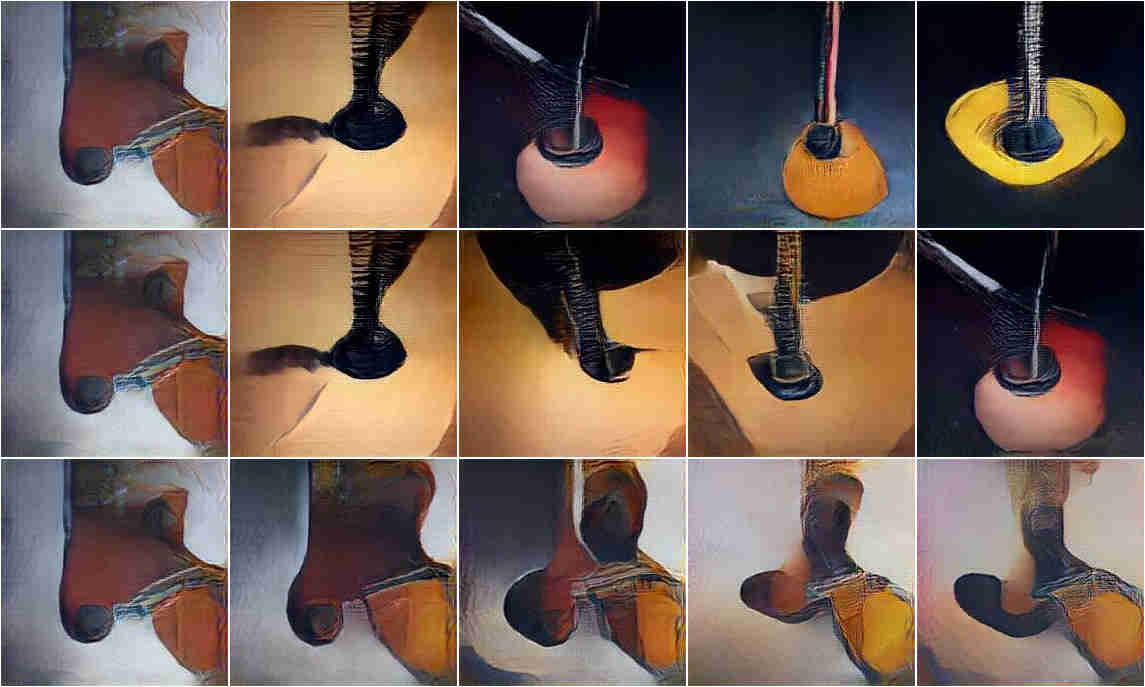}
\end{subfigure}
\caption{$k=5$ randomly chosen images (left) from the MCMC sequence of PPGN
for \emph{Guitar} class, and the corresponding  diverse sets (right) chosen by \CTInherit,
after 20 (bottom), 100 (middle), and 200 (top) MCMC steps.}
\label{fig:guitar}
\end{figure}

\begin{figure}[h]
\centering
\begin{subfigure}{0.45\textwidth}
\centering
\includegraphics[width=0.95\textwidth]{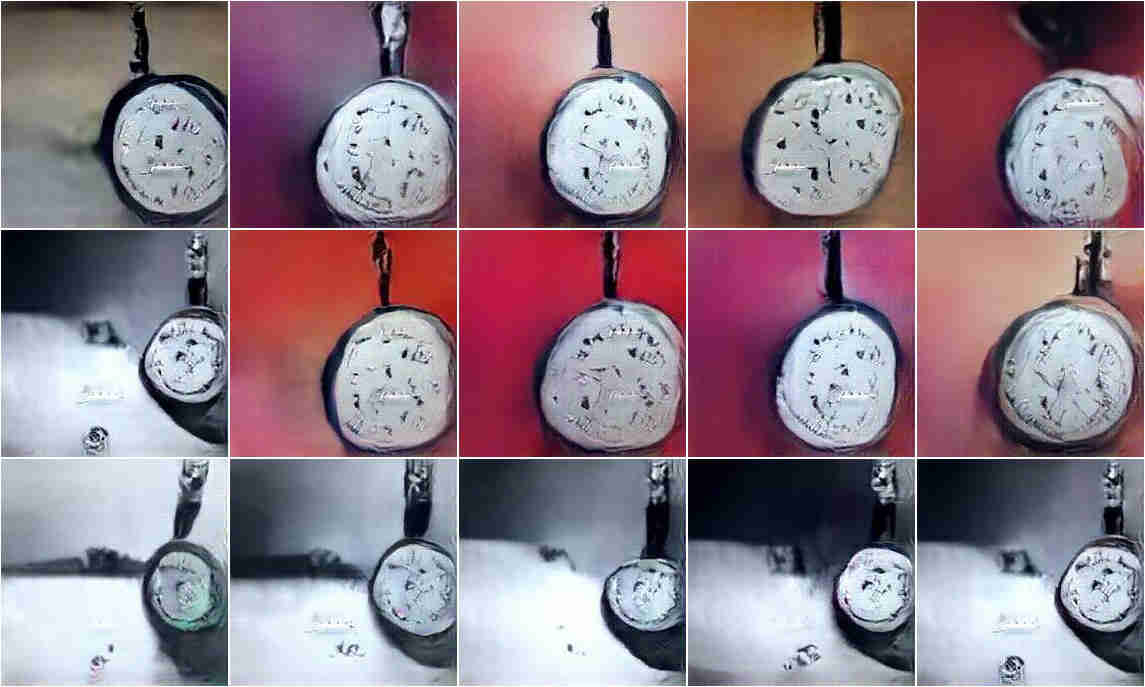}
\end{subfigure}
\centering
\begin{subfigure}{0.45\textwidth}
\includegraphics[width=0.95\textwidth]{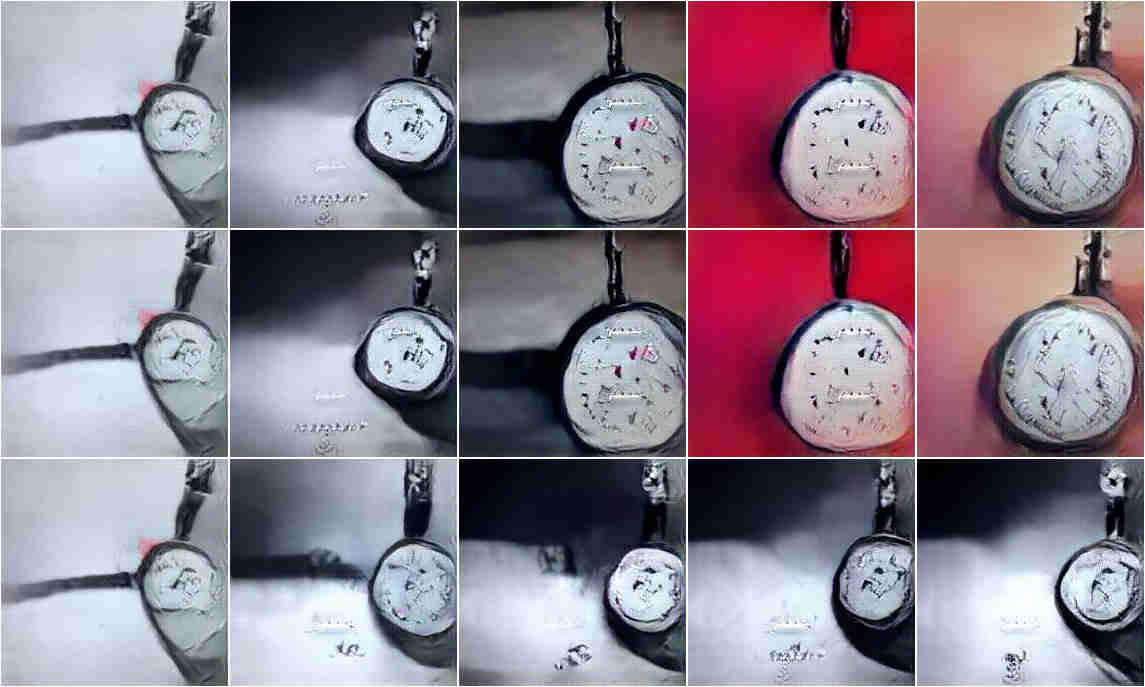}
\end{subfigure}
\caption{$k=5$ randomly chosen images (left) from the MCMC sequence of PPGN
for \emph{Clock} class, and the corresponding  diverse sets (right) chosen by \CTInherit,
after 20 (bottom), 100 (middle), and 200 (top) MCMC steps.}
\label{fig:clock}
\end{figure}

Figs. \ref{fig:greenhouse}--\ref{fig:clock} show the $k=5$ randomly chosen images (left) from the MCMC sequence of PPGN conditioned on several ImageNet classes, and the corresponding  diverse sets (right) chosen by \CTInherit, after 20 (bottom), 100 (middle), and 200 (top) MCMC steps. 
% As the diversity of the image sequence increases with increasing number of MCMC steps, a crucial task for generated image diversification is to select the most diverse subset of these images.
% This can be achieved with \CTInherit. 
% For all examples, the $k=5$ set produced by \CTInherit{}  shows more diverse images compared to random sampling from the corresponding MCMC sequence, as indicated by more diverse shapes (e.g. spiders) and colors (e.g. volcano).
We can see that \CTInherit{} chose more diverse images in shape (e.g., Volcano (Fig.~\ref{fig:vulcano}), Greenhouses, Spiders),
composition (e.g., Sheep, Clocks), 
and color (e.g., Train).

\end{document}